\documentclass[aps,pra,twocolumn,showpacs,nofootinbib]{revtex4-1}
\usepackage {amsmath}
\usepackage {amssymb}
\usepackage[dvips]{graphicx}
\usepackage[T2A]{fontenc}
\usepackage[utf8]{inputenc}
\usepackage{indentfirst}

\usepackage[usenames]{color}
\usepackage{ulem}
\usepackage[colorlinks=true]{hyperref}

\newcommand{\rmi}{\mathrm{i}}
\newcommand{\rme}{\mathrm{e}}
\newcommand{\const}{\mathrm{const}}

\begin{document}
	
\title{MCF solitons and laser pulse self-compression at light bullet excitation in the central core of MCF}
	
\author{A.\,A.\,Balakin}
\author{A.\,G.\,Litvak}
\author{S.\,A.\,Skobelev}
\email{sksa@ufp.appl.sci-nnov.ru}
	
\affiliation{Institute of Applied Physics RAS, 603950 Nizhny Novgorod, Russia}
	
\date{\today}
	
\begin{abstract}
The propagation of laser pulses in multi-core fibers (MCF) made of a central core and an even number of cores located in a ring around it is studied. Approximate quasi-soliton homogeneous solutions of the wave field in the considered MCF are found. The stability of the in-phase soliton distribution is shown analytically and numerically. At low energies, its wave field is distributed over all MCF cores and has a duration, which exceeds the duration of the NSE soliton with the same energy by many (five-six) times. On the contrary, almost all of the radiation at high energies is concentrated in the central core with a duration similar to the NSE soliton. The transition between the two types of distributions is very sharp and occurs at a critical energy, which is weakly dependent on the number of cores and on the coupling coefficient with the central core. The self-compression mechanism of laser pulses was proposed. It consists in injecting such MCF with a wave packet being similar to the found soliton and having an energy larger than the critical value. It is shown that the compression ratio weakly depends on the energy and the number of cores and is approximately equal to 6 times with an energy efficiency of almost 100\%. The use of longer laser pulses allows one to increase the compression ratio  up to 30-40 times with an energy efficiency of more than 50\%. The obtained analytical estimates of the compression ratio and its efficiency are in good agreement with the results of numerical simulation.
\end{abstract}
	
\maketitle
\section{Introduction}\label{sec:1}

One of the current trends in modern fiber optics is associated with the use of micro- and nano-structured systems for the light flux control. In recent years, a whole section of nonlinear science has been formed, which is devoted to theoretical and experimental studies of wave processes in spatially periodic nonlinear media, in which main attention is focused on the following issues: supercontinuum generation \cite{Tran, Panagiotopoulos}, shortening of laser pulse duration \cite{Rubenchik, Aceves, Balakin18, Aceves, Tran14, Cheskis, Skobelev18, Eisenberg2001}, control of the wave field structure \cite{Skobelev2018, Boris, Christodoulides89, Christodoulides2003, Turitsyn12, Lederer2008}, formation of light bullets \cite{Eilenberger2011, Christodoulides2010,Minardi, Mihalache2004, Lobanov2010, Wright, Leblond2017}, soliton-like solutions \cite{Leblond2017, Leblond2016}, and generation of intense laser pulses in active fiber systems.

Studies have shown that self-focusing of wide wave field distribution with a power exceeding a certain critical value \cite{Balakin16} (which differs from the critical self-focusing power in a homogeneous non-linear medium) leads to decomposition of the wave field into a set of incoherent structures \cite{Balakin18} in the process of propagation in a medium with a periodic set of weakly coupled optical fibers. For stable operation with more powerful wave beams, one can use multi-core optical fibers (MCF) with a small number of cores. An example is an MCF, consisting of a central core and an even number of cores located in a ring around it~\cite{Turitsyn12, Skobelev2018, Rubenchik, Rubenchik2013}. Non-uniform stationary nonlinear wave field distributions in such MCFs were found, and their stability was shown even at a total power much higher than the critical self-focusing power~\cite{Skobelev2018}. It is also of interest to study the existence of coherent soliton-like optical pulses in such MCFs, which can propagate along extended paths without changing the structure, in particular, the formation of three-dimensional spatio-temporal solutions that retain their shape due to the balance of diffraction, dispersion of group velocity and nonlinear phase modulations. It is also promising to use self-compression of laser pulses in the MCF using such solutions.

In this paper, we study analytically and numerically the existence, the structure, and the stability of soliton-like wave field distributions in MCF, which is an array of $2N$ identical cores surrounding the central core (Fig.~\ref{ris:ris1}). The main attention is paid to the case of a uniform distribution of the wave field over the ring, which most effectively interacts with the field in the central core. It is shown that radiation is captured into the central core of the MCF when the energy of injected soliton-like laser pulses exceeds some value. This process is accompanied by a significant decrease in the pulse duration. Evaluations of the effectiveness of such self-compression and the achievable minimum pulse duration are confirmed by the results of numerical simulation.

The work is arranged as follows. In Section~\ref{sec:2}, the basic equations are formulated. Section~\ref{sec:3} discusses soliton solutions in the absence of the central core. In Section~\ref{sec:4}, approximate quasi-soliton solutions of the wave field in the considered MCF are found allowing for the central core. Section~\ref{sec:5} analyzes stability of the found solutions. It is shown that only the branch corresponding to the in-phase solitons in the cores is stable. In Sections~\ref{sec:6} and \ref{sec:7}, the method of self-compression of soliton-like and longer laser pulses was proposed and studied. This self-compression leads to formation of a light bullet in an MCF. Elongated laser pulses ensure a significant increase in the compression ratio. Estimates of self-compression effectiveness and the achievable minimum duration of the resulting light bullet are obtained and confirmed by the results of numerical simulation. Section~\ref{sec:8} is devoted to optimizing the parameters of laser pulses to achieve the maximum compression ratio and energy efficiency. In the Conclusion, the main results of the work are formulated.

\section{Basic equations}\label{sec:2}

\begin{figure}
	\includegraphics[width = 0.7\linewidth]{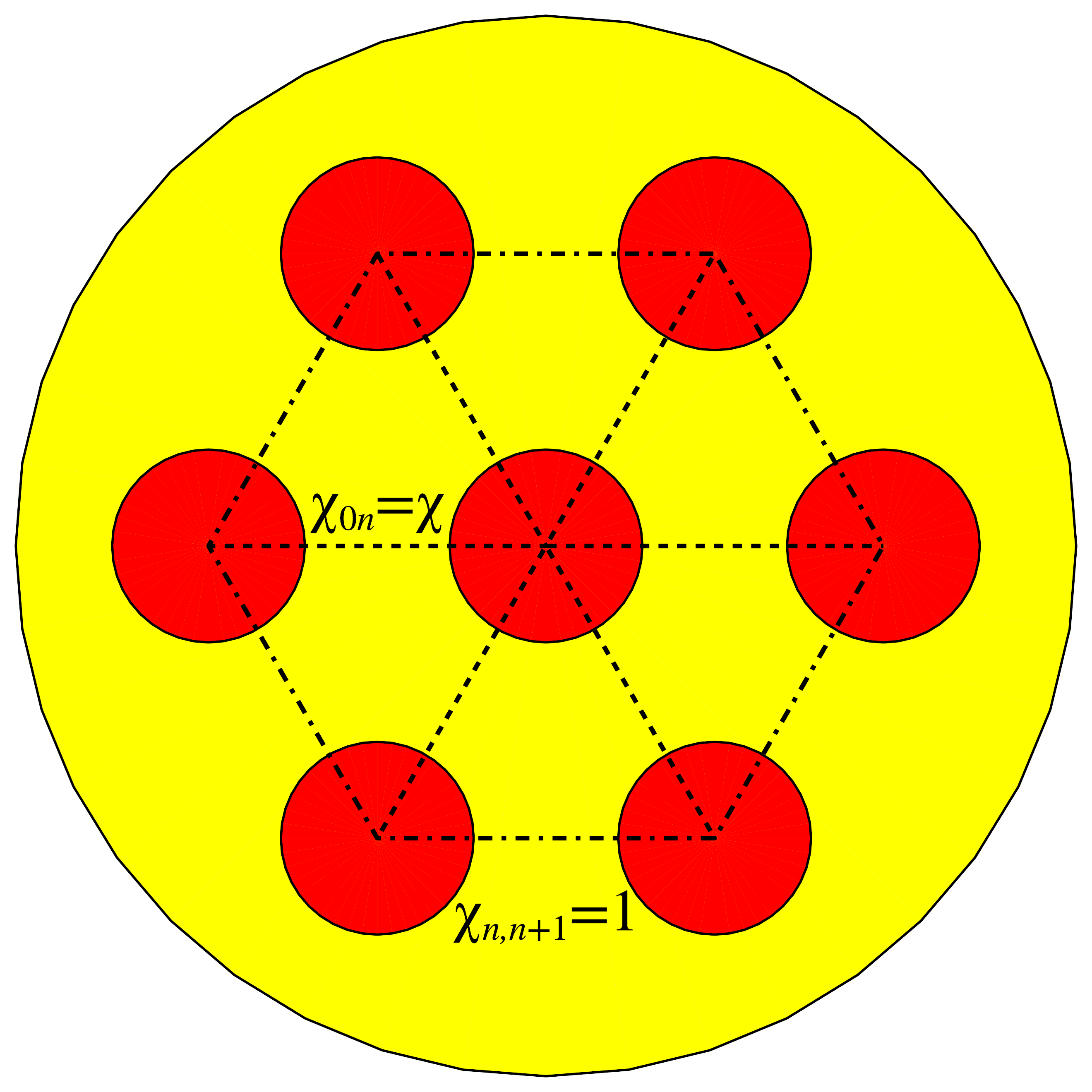}
	\caption{(Color online) Scheme of the MCF under consideration showing the cores arranged in a circle around the central core.
	}\label{ris:ris1}
\end{figure}

Let consider the self-action of wave packets in the MCF, which includes $2N$ identical cores encircling the central one. Figure~\ref{ris:ris1} shows schematically such an MCF with $N = 3$. We will analyze this problem on the basis of the standard theoretical model \cite{Kivshar,Balakin18,Christodoulides89,Balakin16,Aceves,Lederer2008}, within which it is assumed that the fundamental guided modes of optical cores oriented parallel to the $z$ axis are weakly coupled. In this case, the propagation of laser pulses in the MCF can be approximately described as a superposition of fundamental modes localized in each core:
\begin{equation}\label{eq:1}
\mathcal{E}(z,x,y,t) \simeq \sum\limits_{n}\mathcal{A}_n(z,t)\mathcal{F}(x-x_n,y-y_n)e^{\rmi k_n z-\rmi\omega t }+c.c. ,
\end{equation}
where $\mathcal{F}$ is the structure of the fundamental spatial mode in the core, and $\mathcal{A}_n$ is the envelope of the electric field in the $n$th core, which slowly changes along the $z$ axis. The evolution of the envelope in the $n$th core during the propagation of the wave field along the $z$ axis can be influenced by linear dispersion and Kerr nonlinearity of a single core and the interaction with the nearest neighboring cores due to the weak overlapping of the modes guided by them. Assuming that the core coupling is weak and does not perturb the structure of the fundamental mode, we obtain the following system of equations for the envelope of the electric field $\mathcal{A}_n$ in the $n$-th core: 
\begin{multline}\label{eq:2}
\rmi\dfrac{\partial\mathcal{A}_n}{\partial z}+\rmi \dfrac{\partial k_n}{\partial \omega}\dfrac{\partial\mathcal{A}_n}{\partial t}=\dfrac12\dfrac{\partial^2k_n}{\partial \omega^2}\dfrac{\partial^2\mathcal{A}_n}{\partial t^2}+\\+\gamma_n |\mathcal{A}_n|^2\mathcal{A}_n+\sum\limits_{m=0}^{2N}\chi_{mn}\mathcal{A}_m .
\end{multline}
Here, the subscript $n$ varies from $0$ to $2N$, $\gamma_n$ is the nonlinearity coefficient in the $n$th core, the coefficient $\chi_{mn}=\chi_{nm}$ determines the magnitude of the coupling between the $m$-th and $n$-th cores, and $\chi_{nn}=k_n$ is the propagation constant in the cores.

We will assume that all cores are the same, i.e. the propagation constant $k_n \equiv \chi_{nn} = k$, the nonlinearity $\gamma_n = \gamma$, and the coupling coefficients ($\chi_{n,n+1}$, $\chi_{n,0}$ at $n>0$) are the same for all cores. This allows us to write the system of equations \eqref{eq:2} in dimensionless variables: 
\begin{subequations}\label{eq:3}
	\begin{gather}
	\rmi \dfrac{\partial a}{\partial \tilde{z}} = \dfrac{\partial^2a}{\partial\tau^2}+|a|^2a+ \chi \sum\limits_{n=1}^{2N} u_n  ,\\
	\rmi \dfrac{\partial u_n}{\partial \tilde{z}} = \dfrac{\partial^2u_n}{\partial\tau^2}+|u_n|^2 u_n + \chi a + u_{n+1} + u_{n-1} .
	\end{gather}
\end{subequations}
Here, the evolutionary variable $z=\tilde{z}/\chi_{n,n+1}$ is normalized to the coupling coefficient between the cores in the ring, $\tau=\left(t-\frac{\partial k}{\partial\omega}z\right) / \sqrt{\frac12\frac{\partial^2k}{\partial\omega^2} \frac1{\chi_{n,n+1}}}$ is the dimensionless longitudinal variable in the accompanying coordinate system moving with the group velocity of the wave packet; $a\equiv u_0=\rme^{\rmi k z}\mathcal{A}_0 \sqrt{{\gamma}/{\chi_{n,n+1}}}$, $u_n=\rme^{\rmi k z}\mathcal{A}_n \sqrt{{\gamma}/{\chi_{n,n+1}}}$ are the complex amplitudes of the envelope of the wave packet in the central and $n$th cores, respectively. The tilde sign will be omitted in what follows. The parameter $\chi = \chi_{n,0}/\chi_{n,n+1}$ is the normalized coefficient of coupling with the central core.

The applicability of Eqs.~\eqref{eq:3} is limited by the approximation of single-mode propagation of the wave field in each core. It will fail when the radiation power in any core $\mathcal{P}_n=|\mathcal{A}_n|^2\iint|\mathcal{F}|^2dxdy$ becomes close to the critical power of the self-focusing in the environment $\mathcal{P}_{cr}$, i.e. at
\begin{equation}\label{eq:4}
|u_n|^2 \gtrsim \dfrac{4\pi c}{\chi_{n,n+1}\omega_0\iint |\mathcal{F}|^2dxdy} \ggg 1 ,
\end{equation}
where $\omega_0$ is the carrier frequency of the laser pulse, and $c$ is the speed of light. Here, the small factor $\frac{\omega_0}{c} \chi_{n,n+1}\iint |\mathcal{F}|^2dxdy \ll 1$ determines the degree of localization of the fundamental mode on the scale between the cores.

In this paper, we confine ourselves to the study of the simplest case, which corresponds to a uniform distribution of the wave field over the ring ($u_n = f$). In this case, system of equations \eqref{eq:3} takes the form
\begin{subequations}\label{eq:5}
	\begin{gather}
	\rmi\dfrac{\partial a}{\partial z}=\dfrac{\partial^2a}{\partial\tau^2}+|a|^2a+2N\chi f , \label{eq:5a} \\ 
	\rmi\dfrac{\partial f}{\partial z}=\dfrac{\partial^2f}{\partial\tau^2}+|f|^2f +\chi a+2f . \label{eq:5b}
	\end{gather}
\end{subequations}

Eqs.~\eqref{eq:5} conserves the total energy of the wave packet in the process of evolution
\begin{equation}\label{eq:6}
	W=\int\limits_{-\infty}^{+\infty}\left(|a|^2 + 2N |f|^2\right)d\tau= \const . 
\end{equation}
In addition, the presence of the Lagrangian of Eqs.~\eqref{eq:5}
\begin{multline}\label{eq:7}
	\mathcal{L}=\int\limits_{-\infty}^{+\infty} \Bigg\{ \dfrac{\rmi}2 \left(a \dfrac{\partial a^*}{\partial z}-a^* \dfrac{\partial a}{\partial z}\right) - \left| \frac{\partial a}{\partial \tau} \right|^2 + \dfrac12|a|^4 + \\ +
	\rmi N\left(f\dfrac{\partial f^*}{\partial z}-f^*\dfrac{\partial f}{\partial z} \right)-2N\left|\dfrac{\partial f}{\partial\tau}\right|^2+N|f|^4+\\+2N\chi\left(af^*+a^*f\right)+4N|f|^2\Bigg\}d\tau .
\end{multline}
means conserving of the Hamiltonian
\begin{multline*}
H = \int\limits_{-\infty}^{+\infty} \Bigg(\dfrac12|a|^4 + N|f|^4 - \left|\frac{\partial a}{\partial \tau}\right|^2 - 2N \left|\dfrac{\partial f}{\partial\tau}\right|^2+ \\ + 2N\chi\left(af^*+a^*f\right)+4N|f|^2\Bigg)d\tau = \const.
\end{multline*}

\section{Solitons in an MCF without the central core ($\chi=0$)}\label{sec:3}

First, we study the simplest case of the absence of the central core ($\chi = 0$). Then, one can find a soliton-type solution for Eqs.~\eqref{eq:5}
\begin{equation}\label{eq:8}
	u_n = f =\dfrac{\sqrt{2}b\rme^{-\rmi(2+b^2)z}}{\cosh(b\tau)} .
\end{equation}
Unfortunately, this solution will be stable only at small amplitudes. This can be seen from a rough estimate of the field amplitude, at which the solution becomes unstable with respect to azimuthal perturbations. For example, for a wave field with perturbations having the form
\begin{equation}\label{eq:9}
	u_n=\left[f_0+\delta_m\rme^{\rmi\lambda z+\rmi \varkappa_m n} \right]\rme^{-\rmi\left(2+|f_0|^2 \right)z} , ~ |\delta_m|\ll|f_0| ,
\end{equation}
we get real eigenvalues
\begin{equation}\label{eq:10}
	\lambda^2=\left(|f_0|^2-4\sin^2\dfrac{\varkappa_m}{2} \right)^2-|f_0|^4\geq 0
\end{equation}
only for small wave field amplitudes \cite{Skobelev2018}:
\begin{equation}\label{eq:11}
	|f_0|<f_\textrm{cr}=\sqrt{2}\sin\dfrac{\pi}{2N}\mathop{\approx}\limits_{N \gg 1} \dfrac{\pi}{\sqrt{2}N} .
\end{equation}

\begin{figure}
	\includegraphics[width = \linewidth]{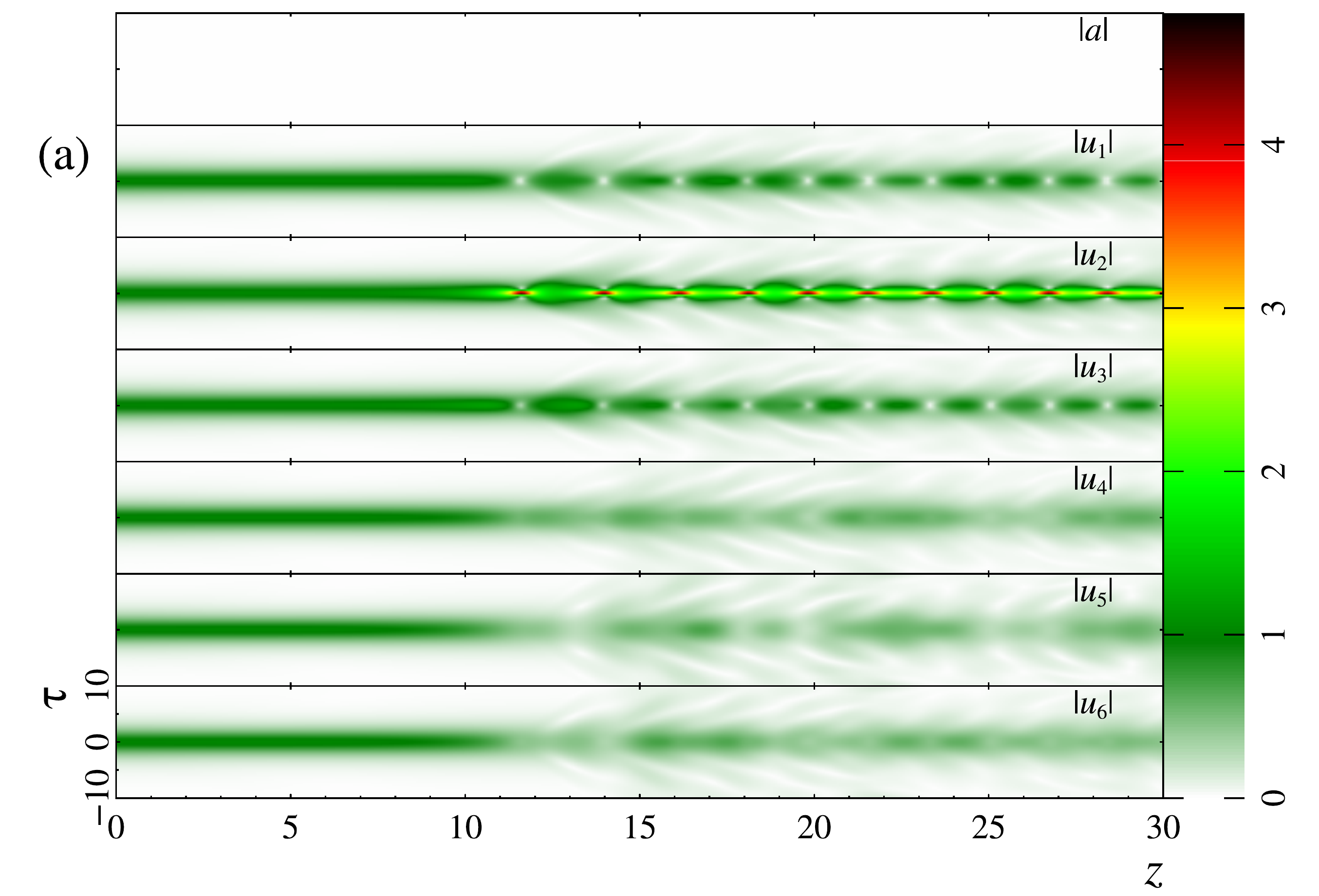}
	\caption{(Color online) Dynamics of the wave field envelope $|u_n|$ in an MCF consisting of six cores ($N = 3$), in the absence of a central core ($\chi = 0$). A laser pulse with initial distribution \eqref{eq:8} with $b = 0.7$ was injected into the fiber.
	}\label{ris:ris2}
\end{figure}

Next, we turn to the results of numerical simulations to confirm the qualitative stability analysis of solution~\eqref{eq:8}. Figure \ref{ris:ris2} shows the dynamics of propagation of a laser pulse with initial distribution~\eqref{eq:8} for the case of $b = 0.7$ in an MCF, which consists of six cores ($N = 3$), in the absence of a central core ($\chi = 0$). The initial wave field amplitude is greater than the critical value ($\sqrt{2}b>f_\textrm{cr}=\sqrt{2}/2$) for these parameters. The figure shows that the instability in this case is quite pronounced. As the wave packet propagates in the medium, radiation is captured in a core (at $z\sim 12$), which further reduces the duration of the laser pulse by several times.

Note, that the introduction of a small modulation of the wave amplitude to the initial distribution of the wave field $u_n=\sqrt{2}b\left[1+10^{-2}\cos(\pi n/3) \right]/\cosh(b\tau)$ allows one to localize radiation in the required core, in contrast to the previously considered case, Eq.~\eqref{eq:8}. Further localization of the radiation could lead to a decrease in the duration of the laser pulse. This mode was studied in~\cite{Rubenchik}. Unfortunately, such regime is rather sensitive to the initial parameters of the wave packet, in particular, to the initial amplitude $b$ and to the amplitude of the perturbations.

Thus, the laser pulse having the form described by Eq.~\eqref{eq:8} and an amplitude greater than critical value \eqref{eq:11} will be unstable with respect to the azimuthal perturbations in the MCF. In this case, the critical amplitude $f_\textrm{cr}$ is not large, and tends to zero as the number of MCF cores increases ($N \to \infty$).

\section{Solitons in an MCF with the central core ($\chi\neq0$)}\label{sec:4}

Let us consider the case, when the central core is present ($\chi \neq 0$), and it becomes possible to transfer energy between the central core and the cores around the MCF. The presence of the central core should provide additional stability with respect to azimuthal perturbations.

\begin{figure}
	\includegraphics[width = \linewidth]{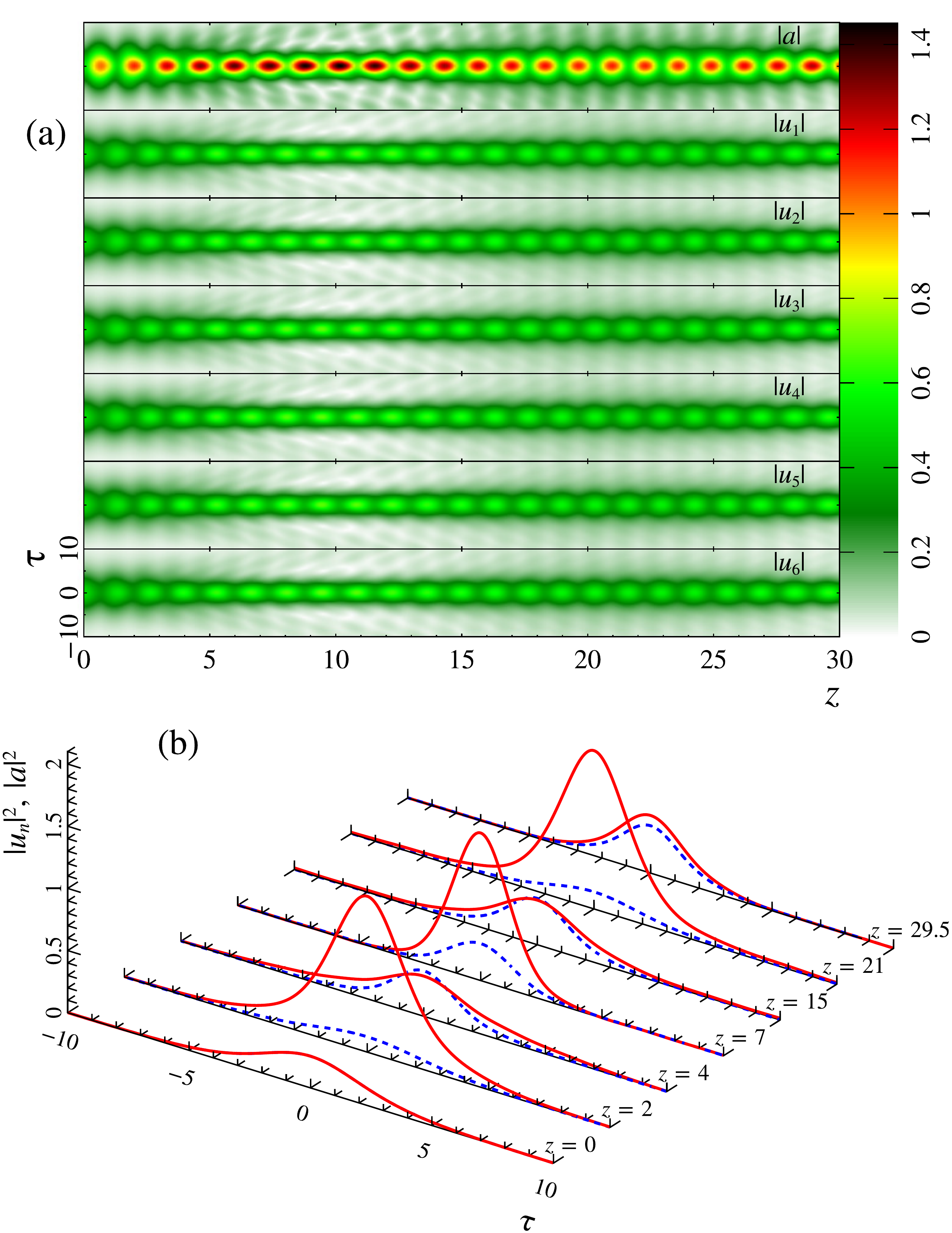}
	\caption{(Color online) \textbf{(a)} Dynamics of the wave field envelope in an MCF consisting of seven cores ($N = 3$) with $\chi = 1$. The initial distribution is \eqref {eq:12}. Fig.~{\bf (b)} shows the intensity distribution of a laser pulse in the central ($|a|^2 $, red lines), and the first ($|u_1|^2$, blue dotted line) cores in different MCF cross sections.
	}\label{ris:ris3}
\end{figure}

Finding analytical soliton solutions of Eqs.~\eqref{eq:5} seems rather difficult. Due to this, at the initial stage, we turn to numerical analysis. Figure \ref{ris:ris3}{\bf(a)} shows the dynamics of the wave field envelope in an MCF consisting of seven cores ($N = 3$), with a coupling coefficient $\chi = 1$. The initial wave field distribution had the following form:
\begin{equation}\label{eq:12}
	a=u_n=\dfrac{\sqrt{2}}{\tau_0\cosh(\tau/\tau_0)} , \quad \tau_0=2.8 .
\end{equation}

It is seen in the figure that the typical dynamics of the laser pulse is associated with the consequent transfer of its energy from the central core to the ring and back. Figure~\ref{ris:ris3}{\bf(b)} shows the dynamics of the wave packet intensity at different points along $z$. It can be seen that the duration of the laser field in the MCF cores also varies slightly obeying a periodic law. Thus, the results of numerical simulation predict the possibility of existence of soliton solutions for Eqs.~\eqref{eq:5}.

We use the variational approach to find an approximate quasi-soliton solution in the MCF. This will allow a more accurate determination of the initial parameters of the laser pulse to eliminate the energy beating between the central core and the ring, in order to eliminate the change in the laser pulse duration in the MCF.

Let us assume that the wave field distribution is close to the form $1/\cosh(\tau)$. This distribution corresponds to the NSE soliton for a single fiber~\eqref{eq:8}. Then, an approximate solution should be sought for in the form
\begin{subequations}\label{eq:13}
	\begin{gather}
	a(z,\tau) = \sqrt{\dfrac{WA}{2\tau_p}} \dfrac{\rme^{\rmi(\phi+\theta+\sigma\tau^2)} }{\cosh(\tau/\tau_p)} , \label{eq:13a} \\
	u_n(z,\tau) = f = \sqrt{\dfrac{W(1-A)}{4N\tau_p}} \dfrac{\rme^{\rmi(\phi+\sigma\tau^2)} }{\cosh(\tau/\tau_p)}  , \label{eq:13b}
	\end{gather}
\end{subequations}
where $W$ is total energy \eqref{eq:6}, $A$ is the fraction of energy in the central core, $\tau_p$ is the pulse duration, $\sigma$ is a parameter of the frequency chirp, and $\theta$ is the relative phase difference between the central core and the ring. Here, we used the presence of the integral of problem \eqref{eq:6}, which is associated with the conservation of the total energy.

Substituting the fields in the form of Eq.~\eqref{eq:13} into expression \eqref{eq:7} and integrating over the variable $\tau$, we get the truncated Lagrangian
\begin{multline}\label{eq:14}
\overline{\mathcal{L}}=W\dfrac{d\phi}{dz}+WA\dfrac{d\theta}{dz}+2\chi W\sqrt{2NA(1-A)}\cos\theta+\\+\dfrac{\pi^2W}{12}\left(\tau_p^2\dfrac{d\sigma}{dz}-4\sigma^2\tau_p^2 \right)+\dfrac{W^2(1-A)^2}{12N\tau_p}+\\+\dfrac{W^2A^2}{6\tau_p}-\dfrac{W}{3\tau_p^2}+2W(1-A) .
\end{multline}
The change in the wave field parameters along the propagation path in the MCF is determined by the Euler equations
\begin{equation}\label{eq:15}
\dfrac{d}{dz}\dfrac{\partial\overline{\mathcal{L}}}{\partial \dot{a}_j}-\dfrac{\partial\overline{\mathcal{L}}}{\partial a_j}=0, \quad \dot{a}_j=\frac{da_j}{dz} .
\end{equation}
Varying Lagrangian \eqref{eq:14}, we arrive at the following system of ordinary differential equations for changing the parameters of wave packet \eqref{eq:13} in the MCF:
\begin{subequations}\label{eq:16}
	\begin{gather}
	\dfrac{dA}{dz}=-2\chi\sqrt{2N}\sqrt{A-A^2}\sin\theta , \label{eq:16a} \\ 
	\dfrac{d\theta}{dz}=\chi\sqrt{2N}\dfrac{2A-1}{\sqrt{A-A^2}}\cos\theta- W\dfrac{(2N+1)A-1}{6N\tau_p}+2 , \label{eq:16b} \\ 
	\dfrac{d\tau_p}{dz}=-4\sigma\tau_p , \label{eq:16c} \\ 
	\dfrac{d\sigma}{dz}=4\sigma^2+\dfrac1{\pi^2\tau_p^3}\left[\dfrac{2NA^2+(A-1)^2}{2N}W-\dfrac4{\tau_p} \right] . \label{eq:16d} 
	\end{gather}
\end{subequations}
The first two equations describe the dynamics of the wave field between the cores, i.e. the change in the fraction of the energy $A$ in the central core and the phase difference $\theta$. The two latter equations determine the dynamics of the pulse duration and the frequency chirp along the MCF. Note that the equation for the common phase $\phi$ splits off from the dynamics of other parameters of the wave packet:
\begin{equation}\label{eq:h}
\frac{d \phi}{dz} = -\frac{4 N (1+\chi)}{2 N+1} - \frac{W^2}{16 (2N+1)^2}.
\end{equation}

Equations~\eqref{eq:16c}, \eqref{eq:16d} have a stationary point of the center type,
\begin{equation}\label{eq:17}
	\tau_p^\textrm{sol}=\dfrac{4}{W}\dfrac1{A^2+(1-A)^2/2N} , \quad \sigma^{sol}=0, 
\end{equation}
which corresponds to soliton-like propagation of a laser pulse. The oscillations near this center are similar to those in the duration of a quasi-soliton pulse in the NSE.

In the stable soliton solution, there are no beats between the wave field in the central core and the isotropic field on the MCF ring. Next, we analyze the possible types of solutions of equations \eqref{eq:16a} and \eqref{eq:16b} taking into account obtained relation \eqref{eq:17} on the phase plane, the form of which essentially depends on the total energy of the wave packet $W$.

\begin{figure}[tb]
	\includegraphics[width = \linewidth]{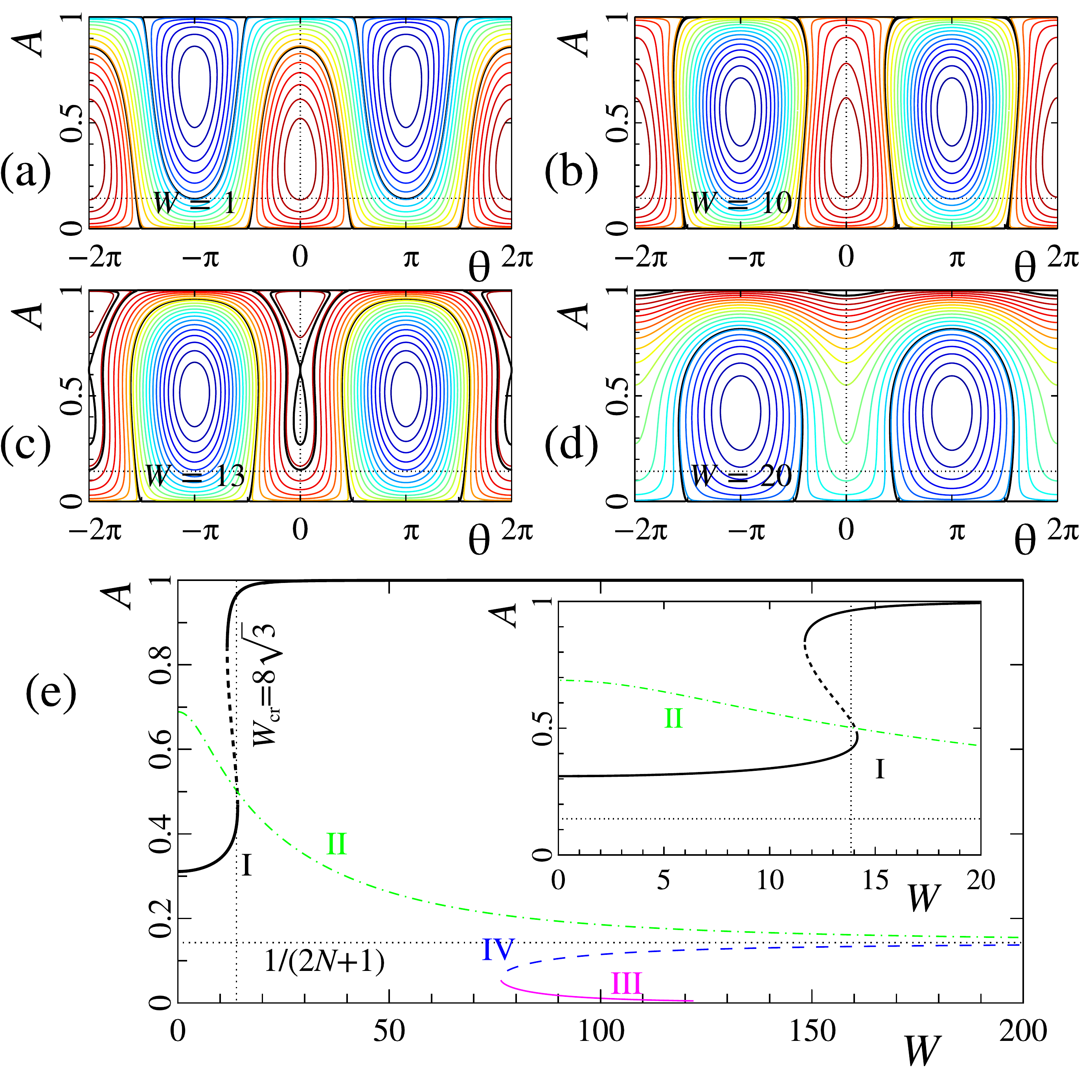}
	\caption{(Color online) {\bf (a-d)} Phase plane of Eqs.~\eqref{eq:16a} and \eqref{eq:16b} allowing for relation \eqref{eq:17} for $N=3$, $\chi=1$ and different values of energy $W$. Bold lines show separatrices. The dotted line corresponds to homogeneous filling $|a|=|f|$. {\bf (e)} Equilibrium states depending on the energy $W$. Dotted lines indicate unstable equilibrium states.} \label{ris:ris4}
\end{figure}

Figures~\ref{ris:ris4}{\bf (a-d)} show the phase plane at various energies. Here, the bold lines show the separatrices. Figure~\ref{ris:ris4}{\bf (e)} presents equilibrium states depending on the energy $W$. As follows from Eq.~\eqref{eq:16a}, there are two distributed equilibrium states of the saddle type ($A = 0$ and $A = 1$), whose positions do not depend on the input energy $W$. These are degenerate one-dimensional manifolds corresponding to the localization of the wave packet of the field only in the central core (for $A = 1$), or only on the MCF ring (for $A = 0$).

The phase plane in the case of a low total energy $W$ of the injected wave packet is presented in Figure~\ref{ris:ris4}{\bf (a)}. It can be seen that system of equations~\eqref{eq:16a}, \eqref{eq:16b} has four equilibrium states: two distributed equilibria of the saddle type ($A = 0$ and $A = 1$) and two centers. With a low energy ($W \to 0$), these two roots $A_\textrm{I}$, $A_\textrm{II}$ are
\begin{subequations}\label{eq:18}
	\begin{gather}
	A^0_\text{I}=\dfrac12-\dfrac1{2\sqrt{2N\chi^2+1}} , \quad \theta=0  , \label{eq:18a} \\
	A^0_\text{II}=\dfrac12+\dfrac1{2\sqrt{2N\chi^2+1}} , \quad \theta=\pi  . \label{eq:18b} 
	\end{gather}
\end{subequations}

As the wave packet energy increases (see Fig.~\ref{ris:ris4}{\bf (b)}), the positions of the centers will shift. The center at $\theta=0$ shifts upwards (branch I in Fig.~\ref{ris:ris4}{\bf (e)}), and the center at $\theta=\pi$ shifts downwards (branch II). It can be seen in Figs.~\ref{ris:ris4}{\bf (a,b)} that the main type of dynamics is the sequential transfer of energy from the central core ($A$ decrease) to the MCF ring and back. Moreover, the beats occur with a significant amplitude if the initial energy fraction in the central core $A$ is not close to the stationary value $A_\text{I}$ or $A_\text{II}$.

With an energy increase to the level
\begin{equation}\label{eq:19}
	W_\textrm{cr} \simeq \frac{8\sqrt{3}}{\sqrt{1-1/(2N)^2}} \approx 8\sqrt{3} \approx 13.86 ,
\end{equation}
a bifurcation occurs, and a new pair of equilibrium states appears: a center and a saddle (see Fig.~\ref{ris:ris4}{\bf (c)}). It is important to note that the critical energy does not depend on the coefficient $\chi$ of coupling with the central core. In the dimensional form (see Eq.~\eqref{eq:3}), the critical energy is proportional to $\sqrt{ \frac{\partial^2k}{\partial\omega^2} \chi_{n,n+1} }/\gamma$, i.e., it can be increased by a denser arrangement of the cores in the ring, which increases the coupling coefficient $\chi_{n,n+1}$. With a further increase in energy, the second bifurcation happens, which is associated with the merging of the lower center with the born saddle. In this case, the born upper center will be pressed to the axis $A \approx 1$. It should be noted that these two bifurcations lead to the appearance of a hysteresis in curve I (see inset in Fig.~\ref{ris:ris4}{\bf (e)}). The range of realization of the hysteresis in the energy is rather narrow, so the change in $A$ has the form of a jump in the case of a small change in $W$. The position of the found hysteresis is shown in branch I as the black dotted line. Finally, there is a third bifurcation at
\begin{equation}\label{eq:20}
W=W_* \approx 12 \sqrt[4]{3}\sqrt{\chi} N^{3/2} + 6\sqrt[4]{3}(5\chi -6\sqrt{3})\sqrt{\chi N},
\end{equation}
which is associated with the birth of two equilibrium states: the center and the saddle, which corresponds to the appearance of the third branch in Fig.~\ref{ris:ris4}{\bf (e)}.

At high energies $W \gg 1$, one can find the asymptotics of all three found branches, which correspond to three equilibrium states of the center type
\begin{subequations}\label{eq:21}
	\begin{gather}
	A_\text{I}\approx 1-\dfrac{288 \chi^2 N}{W^4} , \quad \theta=0 , \label{eq:21a} \\ 
	A_\text{II}\approx \dfrac{1}{2N+1} + \dfrac{24N(2\chi N+2-\chi)}{W^2} , \quad \theta=\pi , \label{eq:21b} \\ 
	A_\text{III}\approx \dfrac{4608\chi^2N^5}{W^4} , ~\theta=0  . \label{eq:21c}
	\end{gather}
\end{subequations}
Note that branch $A_\text{II}$ tends to a uniform intensity distribution $|a|^2 = |f|^2$ that has $A=1/(2N+1)$.

\section{Solution stability analysis}\label{sec:5}

Let us analyze now stability of the found soliton solutions. First, we will focus on the simplest case, which corresponds to branch III. One can see in Fig.~\ref{ris:ris4}{\bf (e)} that this solution exists only at high energies, $W>W_*$. In the case of $N = 3$ and $\chi = 1$, this corresponds to the situation of $W_*\gtrsim 75$. It follows from this figure that the energy fraction in the central core $A$ tends to zero with the increasing energy $W$. This means that almost all of the energy is concentrated in the cores located in the ring. Obviously, this solution is not stable with respect to azimuthal perturbations. This situation is similar to the case of absence of a central core that we considered earlier (Section \ref{sec:3}).

Let us analyze the remaining two branches. The first branch corresponds to the stationary point $\{A_\textrm{I}, \theta=0, \sigma=0, \tau_p\}$, and the second branch, to $\{A_\textrm{II}, \theta=\pi, \sigma=0, \tau_p\}$. The duration $\tau_p$ of the wave structure is determined by expression \eqref{eq:17}. A condition necessary for the stability of the found soliton solution is the presence of a local minimum of the Hamiltonian
\begin{multline}\label{eq:22}
	\overline{H} = -2 \chi W\sqrt{2 N A (1-A)} \cos\theta +\dfrac{\pi^2W}{3} \tau_p^2 \sigma^2 -\\
- \dfrac{W^2}{12N \tau_p} (A-1)^2 - \dfrac{A^2 W^2}{6 \tau_p} + \dfrac{W}{3\tau_p^2}+2AW  .
\end{multline}
This corresponds to positive definiteness of the second derivative of the Hamiltonian in the vicinity of the stationary point. In the case of the Hamiltonian having the form of Eq.~\eqref{eq:22}, the condition of local positive definiteness is noticeably simpler. The reason is that most cross-derivatives on the solution are zero:
\begin{equation}\label{eq:23}
	\dfrac{\partial^2 \overline{H}}{\partial \sigma \partial A} = \dfrac{\partial^2 \overline{H}}{\partial \sigma \partial \theta} = \dfrac{\partial^2 \overline{H}}{\partial \sigma \partial \tau_p} = \dfrac{\partial^2 \overline{H}}{\partial \theta \partial A} = \dfrac{\partial^2 \overline{H}}{\partial \theta \partial \tau_p} =0 .
\end{equation}
Moreover, some of the second derivatives on the solution are always positive.
\begin{subequations}\label{eq:24}
	\begin{gather}
		\dfrac{\partial^2 \overline{H}}{\partial \sigma^2} = \dfrac{2\pi^2}{3} W \tau_p^2 > 0 , \label{eq:24a} \\  \dfrac{\partial^2 \overline{H}}{\partial \tau_p^2} = \dfrac{2 W}{3\tau_p^4} > 0  . \label{eq:24b}
	\end{gather}
\end{subequations}

Thus, a necessary stability criterion is
\begin{subequations}\label{eq:25}
\begin{gather}
	\left(\dfrac{\partial^2 \overline{H}}{\partial A \partial \tau_p}\right)^2 < 4 \dfrac{\partial^2 \overline{H}}{\partial A^2} \dfrac{\partial^2 \overline{H}}{\partial \tau_p^2} , \label{eq:25a} \\ 
	\text{where} \quad \dfrac{\partial^2 \overline{H}}{\partial A \partial \tau_p} = \dfrac{2AN+A-1}{6N} \dfrac{W^2}{\tau_p^2} , \nonumber \\
	\dfrac{\partial^2 \overline{H}}{\partial \theta^2}=2\chi W \sqrt{2NA(1-A)} \cos \theta >0 , \label{eq:25b} \\ 
	 \dfrac{\partial^2 \overline{H}}{\partial A^2}=\dfrac{\chi W  \sqrt{N/2} \cos \theta}{(A-A^2)^{3/2}} - \dfrac{(2N+1)W^2}{6N \tau_p}>0 . \label{eq:25c}
\end{gather}
\end{subequations}
Obviously, the latter conditions, Eqs.~\eqref{eq:25b} and \eqref{eq:25c}, are not fulfilled for mode II, since these second derivatives are negative for $\theta=\pi$. Thus, as follows from the qualitative analysis, branch II is unstable.

\begin{figure}[tb]
	\includegraphics[width = \linewidth]{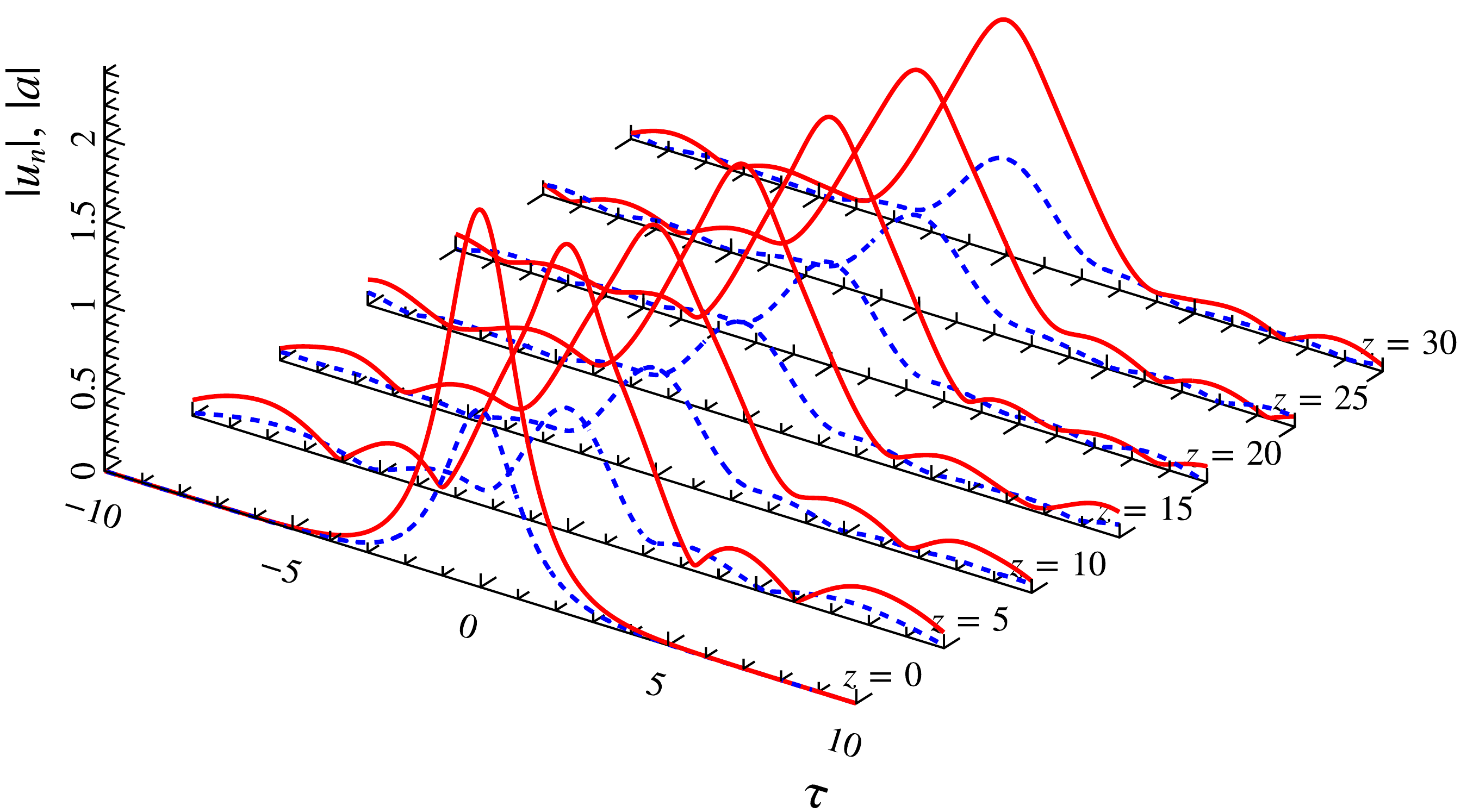}
	\caption{(Color online) Amplitude distributions of the laser pulse in the central $|a(\tau)|$ (solid red line) and in first $|u_1(\tau)|$ (blue dotted line) cores in different MCF cross-sections. A laser pulse \eqref{eq:27} was injected into the MCF input, corresponding to found branch II at $\theta=\pi$, $W=20$. The dispersion length is $L_\textrm{dis}\approx0.2$.} \label{ris:ris5}
\end{figure}

Numerical simulation confirms the instability of branch II. Figure~\ref{ris:ris5} shows the typical dynamics of wave packet \eqref{eq:13}, \eqref{eq:17} in an MCF consisting of seven cores ($N = 3$), with a coupling coefficient $\chi = 1$. A laser pulse with an energy of $W = 20$ was injected into the fiber input
\begin{equation}\label{eq:27}
	a(\tau) = -\dfrac{2.272}{\cosh(1.1982\tau)} , ~u_n(\tau) = \dfrac{1.066}{\cosh(1.1982\tau)}. 
\end{equation}
The minus sign corresponds to the fact that the field in the central core and in the MCF ring are in antiphase. Here, the dispersion length (the length at which the duration of the wave packet has increased by $\sqrt{2}$ times) is $L_\textrm{dis}\approx0.2$. Figure~\ref{ris:ris5} shows that this distribution is unstable in the longitudinal direction. As the wave packet propagates in the MCF, a ``dispersion loss'' of the wave field takes place (formation of a wave ripple extending from the main wave structure), which leads to an increase in the duration of the laser pulse. However, in the process of further evolution of the wave packet, a two-scale quasi-stable distribution of the wave packet is formed (see Fig.~\ref{ris:ris5} for $z>10$). Along with this, the results of numerical simulation demonstrate an important point that the phase difference of the wave field between the central core and the ring varies monotonically in the range $0\leq\theta\leq2\pi$. This means that the coherence of the wave field between the central core and the cores located on the ring is lost.

Let us analyze relations \eqref{eq:25} for the solution that corresponds to branch I. First, we consider the case of low energies $W\ll W_\textrm{cr}$, when the energy fraction in the central core $A$ is close to the value \eqref{eq:18a}. So, the second derivative of $\partial^2\overline{H}/\partial A^2$ \eqref{eq:25c} is positive, since the first term in the expression \eqref{eq:25c} is proportional to the energy $W$ and obviously larger than the second term, which is proportional to $W^3$. It is easy to see that the product of the derivatives $\left(\partial^2\overline{H}/\partial A^2\right) \left(\partial^2\overline{H}/\partial\tau_p^2\right) \propto W^6$ is much larger than the square of the mixed derivative $\left(\partial^2\overline{H}/\partial A\partial\tau_p \right)^2\propto W^8$. Thus, in the case of low energies $W\ll W_\textrm{cr}$, conditions \eqref{eq:25} are satisfied.

For large values of the wave packet energy $W\gg W_\textrm{cr}$, the asymptotic decomposition of the energy share $A$ in the central core \eqref{eq:21a} allows us to make the same conclusion. Indeed, the second derivative $\partial^2\overline{H}/\partial A^2\propto W^7$ is positive, since the first term in expression \eqref{eq:25c} is obviously greater than the second term. Along with this, the product of $\left(\partial^2\overline{H}/\partial A^2\right) \left(\partial^2\overline{H}/\partial\tau_p^2\right) \propto W^{12}$ is much larger than the square of the mixed derivative $\left(\partial^2\overline{H}/\partial A\partial\tau_p \right)^2\propto W^8$. Thus, the performed qualitative analysis demonstrates the stability of branch I. Numerical simulations also prove this stability.

\section{Self-compression of quasi-soliton pulses}\label{sec:6}

A distinctive feature of the found stable soliton solution (branch I) is the presence of hysteresis in the vicinity of $W \approx W_\text{cr}$. In the inset in Fig.~\ref{ris:ris4}{\bf(e)}, the hysteresis-related transition is shown by a vertical dashed line. At low energies ($W<W_\textrm{cr}$), the energy fraction $A$ in the central core is almost constant, $A \approx A_\textrm{I}^0 = 1/2-1/(2\sqrt{2N+1})$. This corresponds to the fact that the wave field of the found solution is distributed quasi-uniformly over all MCF cores. However, at high energies ($W>W_\textrm{cr}$), the wave field is concentrated mainly in the central core ($A\approx 1$) and has the form of a ``light bullet'' with duration \eqref{eq:17} close to the duration of the NSE soliton of the same energy ($\tau_p^\text{NSE}=4/W$). Otherwise ($W<W_\textrm{cr}$), the duration of the found solution turns out to be 5--6 times longer than the duration of the NSE soliton of the same energy. Note that the energy fraction in the central fiber is almost constant in the found solution. This allows us to fix its value at $A_\text{in} = 1/2-1/(2\sqrt{2N+1})$. The jump between the two parts of branch I at $W \approx W_\text{cr}$ allows us to suggest a method of self-compression of a laser pulse with an energy greater than the critical one, $W_\text{cr}$, when laser radiation is captured in the central core (forming a stable distribution with $A \approx 1$ in form of light bullets).

The transition from one part of branch I to another occurs almost without a loss of energy in the hysteresis region $W \approx W_\textrm{cr}$. It yields the compression ratio (the ratio of the initial $\tau_p$ and final $\tau_p^\text{out}$ durations) of a laser pulse as $Q = 1/\big(A_\text{in}^2+{(1-A_\text{in})^2}/{2N} \big)$. A further increase in the initial wave packet energy $W$ at the fixed value of $A=A_\text{in}$ will divert the initial wave packet distribution further from the hysteresis region. Accordingly, the capture of radiation in the ``light bullet'' mode becomes possible for an ever smaller fraction of the energy $\eta<1$.

To capture laser radiation effectively in the central core, it is necessary to get into a small neighborhood of a stable equilibrium state (Fig.~\ref{ris:ris4}) with the parameters $A \approx 1$ and $|\theta| \lesssim 1$. An increase in the fraction of the energy $A$ is obtained automatically, when the central core interacts with a large number of cores in the ring. The second condition is more difficult to reach. Indeed, assuming that the length of the energy transfer to the center is of the order of magnitude of $1/2 \chi$, we can roughly estimate the change in $\delta \theta$ from Eqs.~\eqref{eq:16b} and \eqref{eq:17} for $A \approx A_\text{in}$:
$$|\delta \theta| \approx \frac{1}{2\chi} \frac{W^2}{48 N^2} \big[(2N+1)A_\text{in}-1\big] \big[2NA_\text{in}^2+(1-A_\text{in})^2 \big] \le 1.$$
Thus, the proximity condition to the desired equilibrium state $|\delta \theta| \le 1$ can only be fulfilled for energies being lower than
\begin{equation}\label{eq:wmax}
W_\text{lim} \simeq \frac{4N \sqrt{6\chi}}{\sqrt{[(2N+1)A_\text{in}-1] [2NA_\text{in}^2+(1-A_\text{in})^2 ]}}.
\end{equation}
For the considered parameters $\chi = 1$, $N = 3$ and $A_\text{in} = 0.31$, the maximal energy is equal to
\begin{equation}\label{eq:wmax_N3}
W \le W_\text{lim} \approx 26 \quad \Rightarrow \quad \tau_p^\text{out} \ge 4/W_\text{lim} \approx 0.15.
\end{equation}
Consequently, the energy of the light bullet generated in the process of self-compression is limited by the energy value $W_\text{lim} \approx 26$, and the duration of the output pulse will be longer than 0.15.

The idea of the minimum duration and, accordingly, the maximum energy of the resulting light bullet allows us to predict the dynamics of the wave field with a further increase in the initial energy. Indeed, as soon as the difference between the initial $W$ energy and $W_\text{lim}$ exceeds the critical energy $W_\text{cr}$, the system will be able to form \emph{several} diverging, less energetic pulses. In other words, in the process of capturing radiation into the central core, the laser pulse will be split up in the longitudinal direction into several wave structures for energies
\begin{equation} \label{eq:w_bound}
W \gtrsim W_\text{lim}+W_\text{cr} \approx 40 .
\end{equation}

Let us estimate the compression efficiency $\eta$. It is obviously close to 100\% ($\eta \simeq 1$ at $W = W_\text{cr}$) near the hysteresis on the phase plane (Fig.~\ref{ris:ris4}). Suppose that the efficiency changes continuously with an increasing total energy, and at the total energy $W \simeq W_\text{lim} + W_\text{cr}$, in the process of splitting, two wave structures arise with an energy being of the order of the critical value, i.e., $\eta \simeq 2 W_\text{cr}/(W_\text{lim}+W_\text{cr})$ at $W \simeq W_\text{lim}+W_\text{cr}$. Then, the simplest approximation for the efficiency is
\begin{equation}\label{eq:eta_appr}
\eta_\text{est} \approx \frac{2 W_\text{cr}}{W+W_\text{cr}}.
\end{equation}
This estimate is most suitable for the pulses, which are close to the solutions in the hysteresis region. In any case, the compression efficiency will be at least
\begin{equation}\label{eq:eta_min}
\eta_\text{min} \approx \frac{W_\text{cr}}{W},
\end{equation}
which corresponds to the formation of a light bullet with an energy equal to the critical value $W_\text{cr}$.

All this allows us to estimate the compression ratio $Q$ as the ratio of the initial $\tau_p$ and the final $\tau_p^\text{out}$ durations
\begin{equation}\label{eq:32_s}
Q = \frac{\tau_p}{\tau_p^\textrm{out}} \approx \frac{1}{4} \eta_\text{est} W \tau_p = \frac{\eta_\text{est}}{ A_\textrm{in}^2+{(1-A_\textrm{in})^2}/{2N} },
\end{equation}
where the final duration $\tau_p^\text{out} \approx 4/\eta W$ is determined by expression \eqref{eq:17} with $A = 1$ and the energy $\eta W$, the initial duration $\tau_p$ is related to the energy of $W$ by the expression \eqref{eq:17} with $A = A_\text{in}$. The compression cannot exceed $Q_\text{max} = 1/(A_\text{in}^2 + (1-A_\text{in})^2/2N)$ obtained at the 100\% light bullet excitation efficiency ($\eta \approx 1$). In the case of $N = 3$, this yields $Q_\text{max} \approx 5.68$.

\begin{figure}
	\includegraphics[width = 0.9 \linewidth]{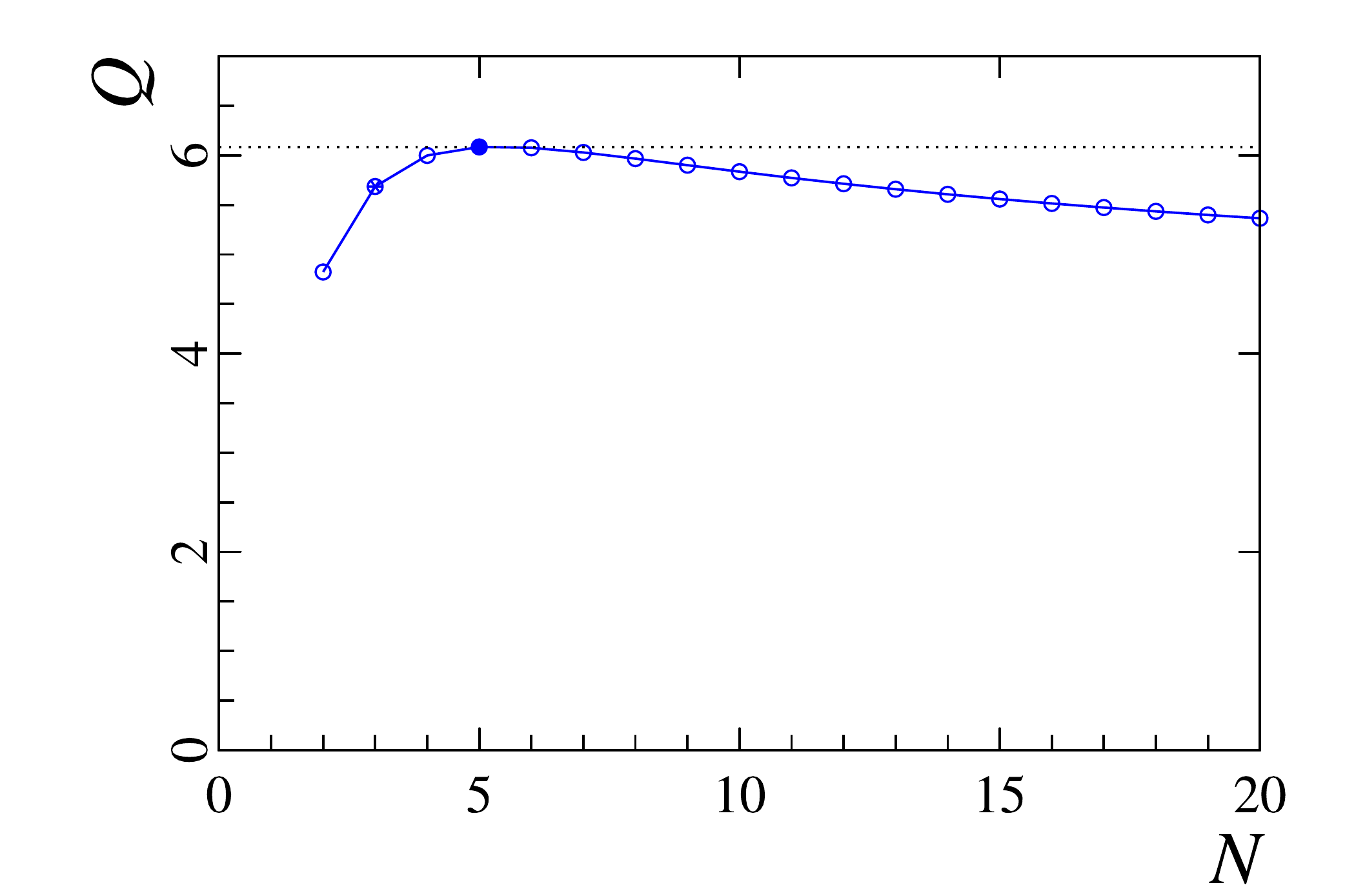}
	\caption{Compression ratio calculated by expression \eqref{eq:32_s} with $\eta_\text{est}=1$ and $A=A_\text{in}$. } \label{ris:ris8}
\end{figure}

So, the use of pulses with the energy $W \simeq W_\text{cr}$ allows one to compress the pulse by 5.68 times with an energy efficiency of 100\% in a seven-core MCF ($N = 3$). Unfortunately, a change in the number of MCF cores leads only to an insignificant change in the compression ratio for $A=A_\text{in}$ (Fig.~\ref{ris:ris8}). It follows from the figure that the compression ratio, in fact, does not depend on the number of cores $2N+1$. Note that the local maximum is reached at $N = 5$ (11 cores). At the same time, an increase in the compression ratio at the maximum is only 6\% as compared with the case of $N = 3$. An increase in the pulse energy while maintaining the value of the product $W \tau_p$ (laser pulses of the soliton form) also leads only to a decrease in the compression ratio due to a decrease in the energy efficiency $\eta$ with increasing energy \eqref{eq:eta_appr}.

\begin{figure}
	\includegraphics[width = 0.99\linewidth]{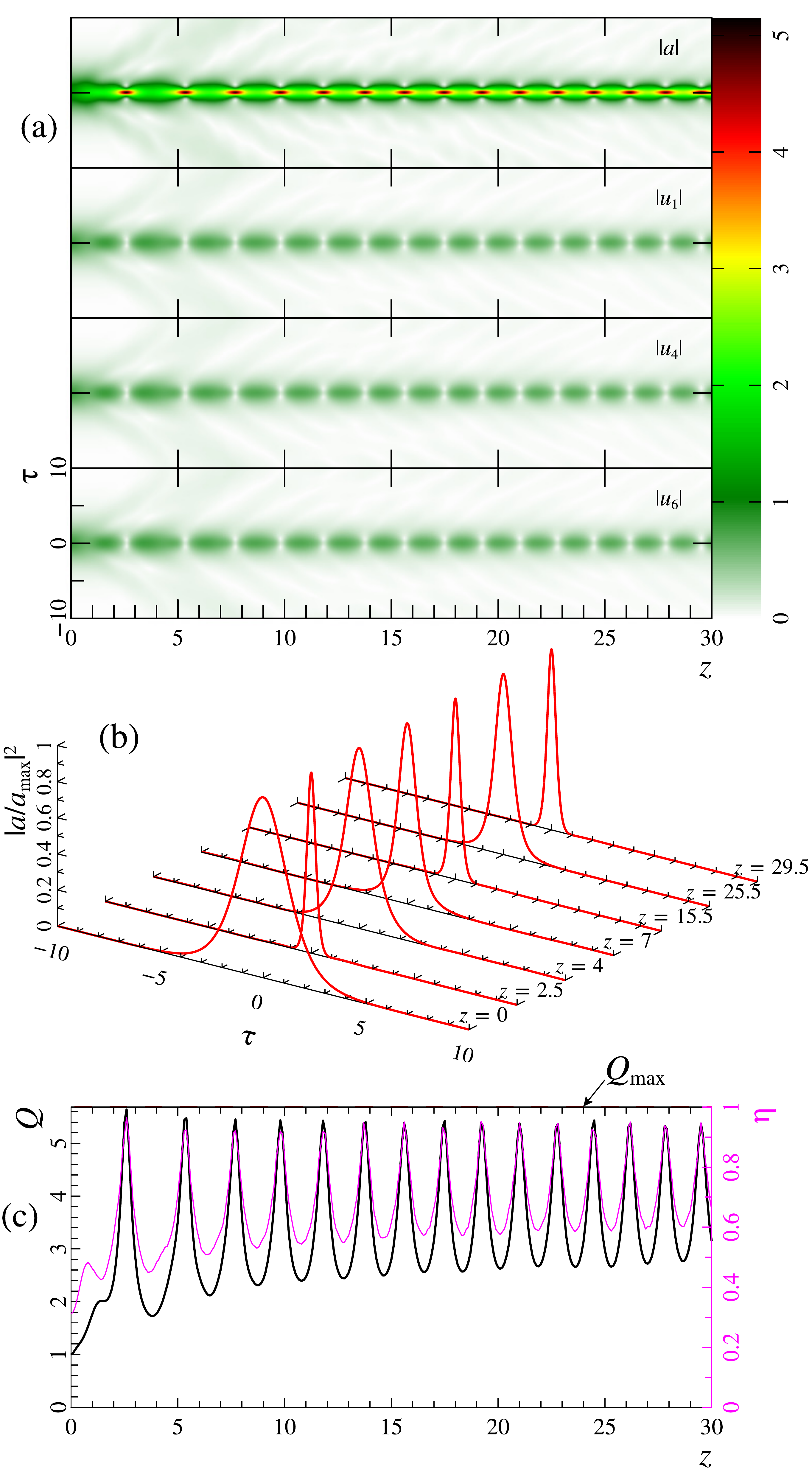}
	\caption{(Color online) {\bf (a,b)} Dynamics of the wave field envelope in an MCF consisting of seven cores ($N = 3$, $\chi = 1$) for the energy $W = 15$. Panel~{\bf(b)} shows the intensity distribution of a laser pulse in the central core $|a|^2$ at different $z$. Panel~{\bf(c)} shows the dependences of the compression ratio $Q$ (thick black line) and the fraction of energy $\eta$ in the central core (thin magenta line) on the $z$ coordinate. The axis for $Q$ is normalized to the maximum value \eqref{eq:32_s} with $\eta_\text{est} = 1$.} \label{ris:ris7}
\end{figure}

The results of numerical simulations of wave packet capturing in the central core are in good agreement with the estimates presented above. In all the calculations presented below, the initial energy fraction $A_\text{in} = A_I^0$ in the central core is equal to that in the linear case (Eq.~\eqref{eq:18a}).

First of all, we consider a laser pulse with the energy $W=15 \gtrsim W_\text{cr}$ and the duration determined by formula \eqref{eq:17}:
\begin{equation}\label{eq:30}
	a(\tau) = \dfrac{1.24}{\cosh(0.66\tau)} , ~u_n(\tau) = \dfrac{0.75}{\cosh(0.66\tau)} . 
\end{equation}
Figure~\ref{ris:ris7} shows that the wave field transforms to a distribution localized predominantly in the central core for the path $z \approx 3$. The magenta line in Figure~\ref{ris:ris7}{\bf (c)} shows the dependence of the fraction of the energy in the central core on the evolutionary variable $z$. It is seen that the energy fraction in the central core increases from $0.31$ to a value comparable with $1$ in the process of transformation of the initial distribution. The black line shows the dependence of the wave packet compression ratio $Q = \tau_p^\text{in}/\tau_p^\text{out}$ on the coordinate $z$. It can be seen that this process is accompanied by a decrease in the duration of the wave packet by about $5.6$ times, which is consistent with the above qualitative analysis. The initial dispersion length of $L_\text{dis} = 0.8$ is significantly less than the length $z_\text{col}$ of the capture in the central core. This means that the considered process is adiabatic.

Figure~\ref{ris:ris7}{\bf (a)} shows the presence of radiated emission (dispersion loss) along the fiber in the process of wave packet capturing in the central core. This is due to the fact that the value of the energy fraction in the central core is set to $A_\text{in} = 0.31$ in the numerical simulations, which is slightly different from the exact value (see Fig.~\ref{ris:ris4}{\bf (e)}). As follows from Figure \ref{ris:ris7}{\bf (c)}, the energy in the central core is comparable with the critical value $W_\text{cr}$. Analysis of the results of the numerical simulation shows that the formed structure does not have a time chirp. It is natural to call this space-time nonlinear structure a \emph{light bullet}.

The observed oscillations in the evolution of the compression ratio $Q$ and the energy fraction $\eta$ in the formed light bullet are mainly associated with the periodic transfer of energy between the central core and the MCF ring. These energy beats are due to the fact that the parameters of $A$ and $\theta$ for the formed wave structure differ slightly from the exact values of the center on the phase plane (see Fig.~\ref{ris:ris4}\textbf{(e)}).

\begin{figure}
	\includegraphics[width = \linewidth]{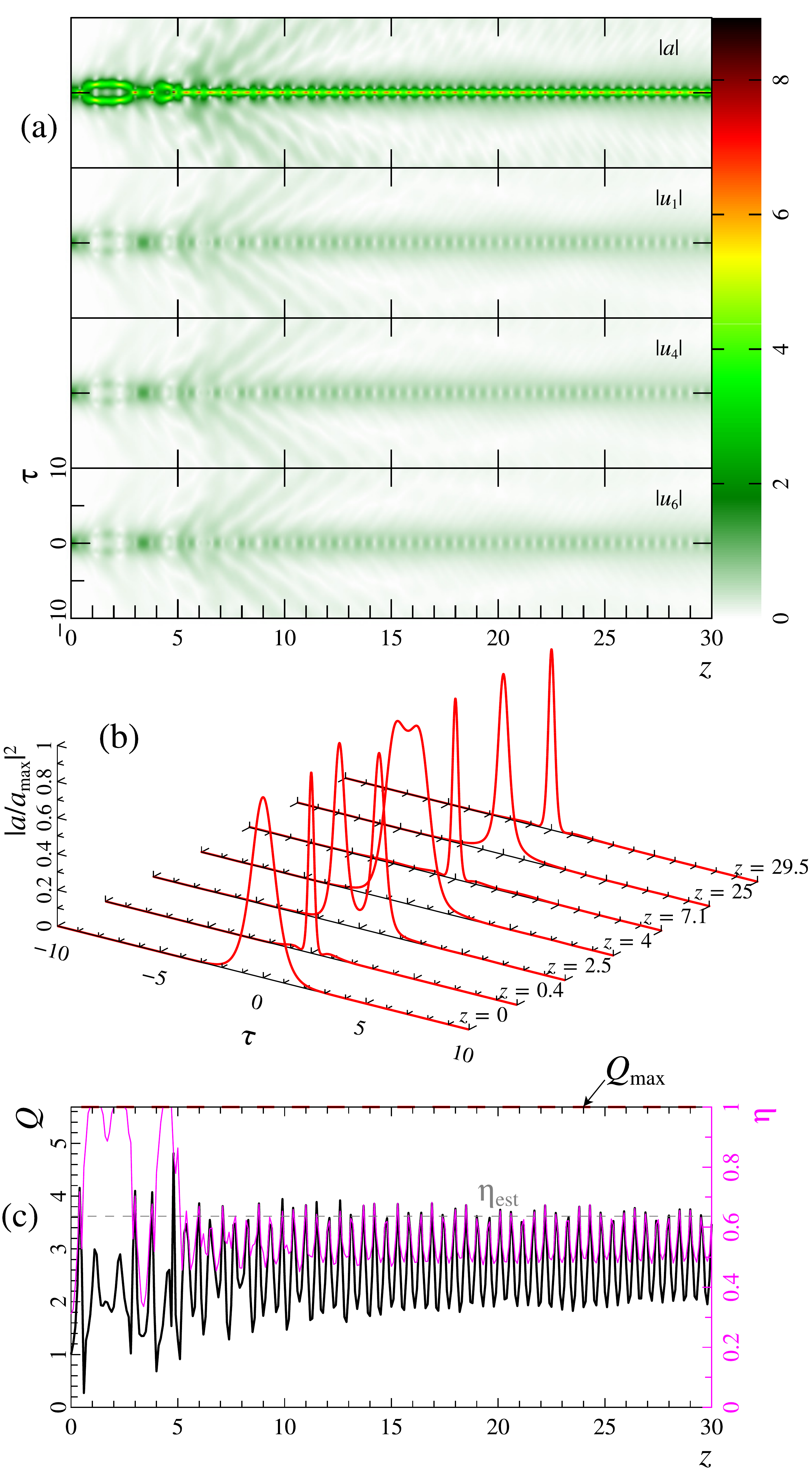}
	\caption{(Color online) {\bf (a,b)} Dynamics of the wave field envelope in 7-core MCF ($N = 3$, $\chi = 1$) for the energy $W = 30$. The initial distribution is determined by the Eq.~\eqref{eq:34}. Captions for the figure are the same as in Fig.~\ref{ris:ris7}. The dotted line in Fig.~\textbf{(c)} shows the efficiency estimated by Eq.~\eqref{eq:eta_appr}.
	} \label{ris:ris10}
\end{figure}

A twofold increase in the energy of a laser pulse leads to a stronger dispersion loss and, accordingly, to a decrease in the compression ratio. Figure~\ref{ris:ris10} shows the dynamics of a wave packet with the energy $W = 30$ and the duration according to Eq.~\eqref{eq:17}:
\begin{equation}\label{eq:34}
	a(\tau) = \dfrac{2.48}{\cosh(1.32\tau)} , ~u_n(\tau) = \dfrac{1.51}{\cosh(1.32\tau)} . 
\end{equation}
One can see a much more complex and prolonged (up to $z \approx 7$) dynamics of light bullet formation. Moreover, this process is accompanied by a noticeable radiation emission in the longitudinal direction (oblique ``tails'' in Figure~\ref{ris:ris10}\textbf{(a)}). At the same time, efficiency estimate \eqref{eq:eta_appr} shows good agreement with the results of the numerical simulation (the dotted line in Fig.~\ref{ris:ris10}\textbf{(c)}). In this case, the compression ratio turned out to be $Q \approx 3.6 < Q_\text{max} \approx 5.7$ with efficiency of about 63\%. Thus, an increase in the energy of the wave packet with preservation of the soliton pulse shape (according to \eqref{eq:17}) leads to deterioration of all compression parameters as the distance from the critical energy decreases. This agrees well with the previous qualitative analysis.

\section{Self-compression of long pulses}	\label{sec:7}

\begin{figure}
	\includegraphics[width = \linewidth]{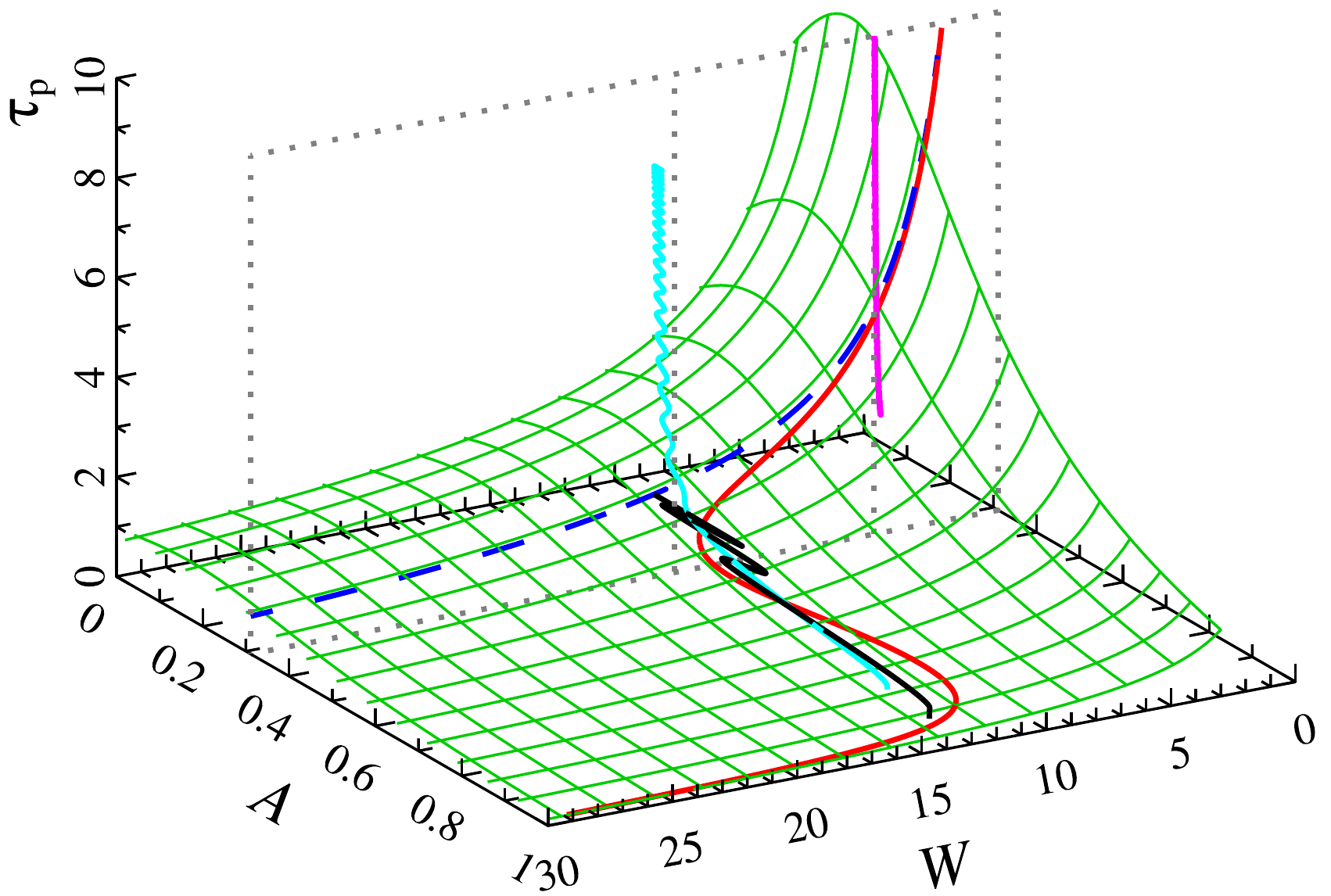}
	\caption{(Color online) The green surface is the dependence \eqref{eq:17} of the soliton duration $\tau_p^\text{sol}$ on $W$ and $A$; the red curve corresponds to the stationary values of $\{\tau_p^\text{sol}, A_I\}$ on the stable branch $I$ (Fig.~\ref{ris:ris4}{\bf (e)}); the blue dash is the duration $\tau_p^\text {sol}$ at $A = A_\text{in}$. The remaining curves show the dynamics of the wave packet parameters within Eqs.~\eqref{eq:16} with the initial $A = A_\text{in}$ for three cases: (i) black line for $\tau_p = \tau_p^\text{sol}$ and $W \approx W_\text{cr}$; (ii) blue line for $\tau_p = 5 \tau_p^\text{sol}$ and $W \approx W_\text{cr}$; (iii) magenta line for $\tau_p = 5 \tau_p^\text{sol}$ and $W = 5 < W_\text{cr}$.
	} \label{ris:ris15}
\end{figure}


So, it is not possible to obtain output pulses shorter than $4/W_\text{lim}$ in the process of radiation capturing into the central core. However, we can get the same pulse from an initially longer laser pulses and thereby significantly increase the compression ratio. Increasing neither the number of cores, nor the energy of a soliton-like wave packet allows one to achieve a compression ratio greater than 6. Therefore, the only possibility to increase the compression ratio $Q$ is to increase the initial duration of the wave packet $\tau_p$ at a fixed total energy, i.e., using the wave packet in the form of \eqref{eq:13} without explicit relation \eqref{eq:17} of duration and energy at the equilibrium state. 

Figure~\ref{ris:ris15} shows the possibility of a significant increase in the compression ratio $Q$ of a laser pulse compared to $Q_\text{max} \approx 6$. In the figure, the green surface shows dependence \eqref{eq:17} of the duration $\tau_p^\text{sol}$ of the found solution on the energy $W$ and the fraction of the energy in the central core $A$. The red line shows the equilibrium states of $\{\tau_p^\text{sol}, A_I \}$ at the stable branch $I$ depending on the energy $W$ (see Fig.~\ref{ris:ris4}{\bf (e)}). Also, Figure \ref{ris:ris15} shows the trajectories of the wave packet parameters $\{\tau_p, A\}$ calculated in the framework of Eqs.~\eqref{eq:16} for three cases. The black line corresponds to the wave packet with the duration equal to the found solution $\tau_\text{in} = \tau_p^\text{sol}$ at the energy $W \approx W_\text{cr}$ and $A = A_\text{in}$. It can be seen from the figure that the duration of the wave packet decreases as the radiation is captured in the central core ($A \to 1$). In this case, the trajectory is trapped in the vicinity of equilibrium state.

The use of longer initial pulses (five times longer for the cyan curve in Fig.~\eqref{ris:ris15}) allows us to distinguish two stages in the dynamics. Initially, the nonlinearity significantly exceeds the media dispersion, and the duration of the wave packet decreases obeying the law $d^2 \tau_p / dz^2 \propto -W / \tau_p^2$. In this case, the fraction of the energy in the central core almost does not change $A \approx A_\text{in}$. At that moment, when the wave packet duration becomes approximately equal to the value of the equilibrium state~(Eq.~\eqref{eq:17}) (red line in Fig.~\ref{ris:ris15}), the capture is taking place in the central core and accompanied by a decrease in the wave packet duration similar to the black curve. It can be seen from the figure that the trajectory of $\tau_p, A$ (cyan line) turns out to be in the vicinity of the equilibrium state (red line). In this case, the generalization of the expression \eqref{eq:32_s} for the compression ratio in this case is trivial:
\begin{equation}\label{eq:32}
Q = \frac{\tau_p}{\tau_p^\textrm{out}} \approx \frac{1}{4} \eta_\text{est} W \tau_p \equiv \frac{\eta_\text{est} W/W_0}{ A_\textrm{in}^2+{(1-A_\textrm{in})^2}/{2N} },
\end{equation}
Here, we introduce the definition of the energy 
$$ W_0 = \frac{4}{\tau_p} \frac{1}{A_\text{in}^2 + (1-A_\text{in})^2/2N} \le W,$$
which characterizes the initial duration of $\tau_p$ in accordance with formula \eqref{eq:17}.

It should be noted that the use of long pulses with an energy significantly lower than the critical one ($W = 5 <W_\text{cr}$ for the magenta curve in Fig.~\ref{ris:ris15}) leads to oscillations in the duration near the equilibrium state, similar to the oscillations of the quasi-soliton solution for the NSE. The reason is the absence of a stable solution for such energies corresponding to the capture in one core.

\begin{figure}
	\includegraphics[width = \linewidth]{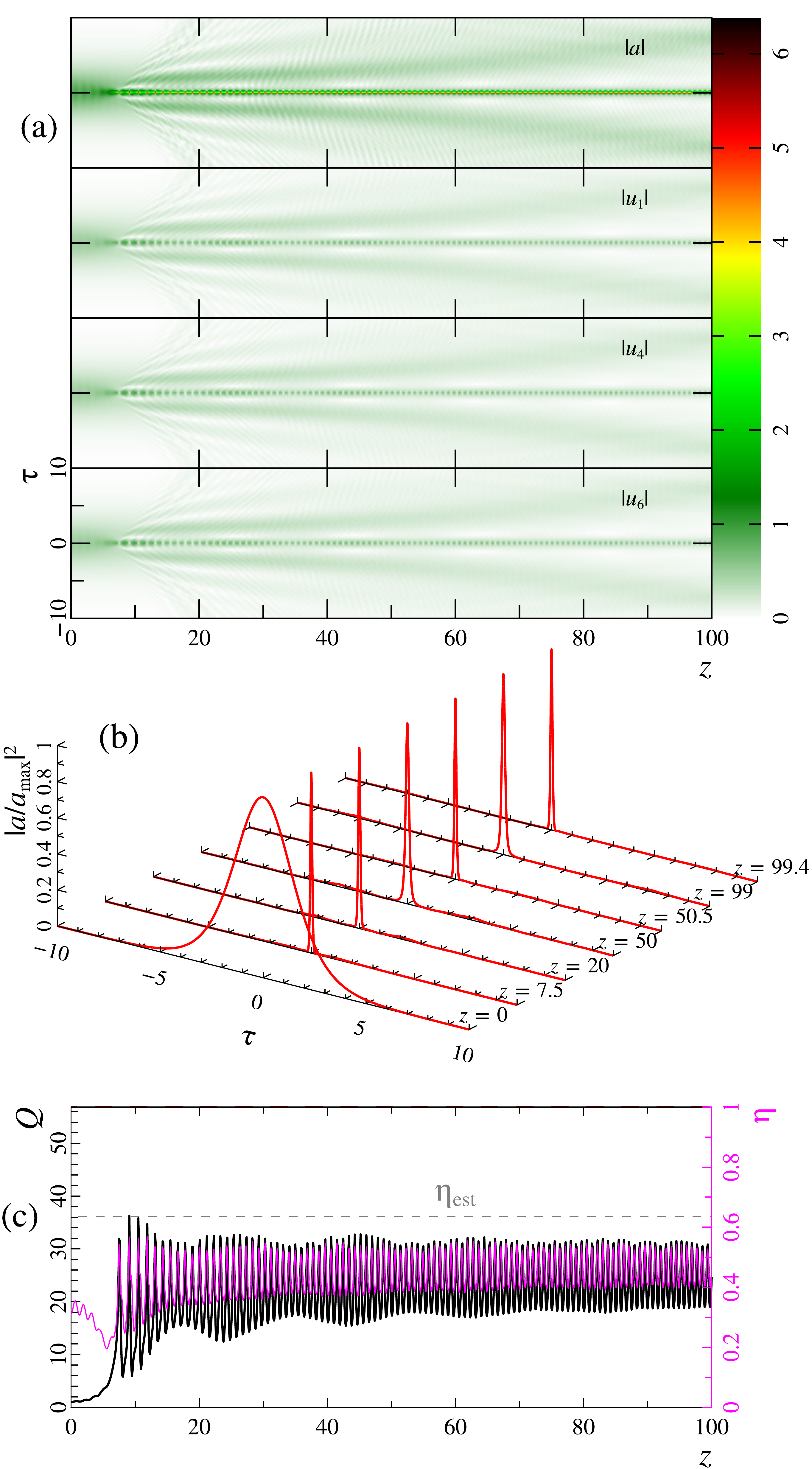}
	\caption{(Color online) {\bf (a,b)} Dynamics of the wave field envelope in a 7-core MCF ($N = 3$, $\chi = 1$) for the energy $W = 30$ and the duration $\tau_p=7.58$. The initial distribution is determined by Eq.~\eqref{eq:W0_3}. Captions for the figure are the same as in Fig.~\ref{ris:ris7}. The dotted line in Fig.~\textbf{(c)} shows the efficiency estimated by Eq.~\eqref{eq:eta_appr}. The axis for $Q$ is normalized with respect to the maximum value \eqref{eq:32} with $\eta_\text{est} = 1$.} \label{ris:ris14}
\end{figure}

Numerical simulation confirms that the use of longer laser pulses can significantly increase the compression ratio at the expense of a slight decrease in the energy efficiency. Figure~\ref{ris:ris14} demonstrates this through the example of a wave packet with the initial energy $W = 30$ and the duration $\tau_p = 7.58$ (corresponding to $W_0 = 3$):
\begin{equation}\label{eq:W0_3}
a(\tau) = \dfrac{0.784}{\cosh(0.132\tau)} , ~u_n(\tau) = \dfrac{0.477}{\cosh(0.132\tau)} . 
\end{equation}
The figure shows three characteristic stages of the wave packet evolution. At the first stage ($z \lesssim 6$), the dynamics of the wave packet is close to the periodic evolution of a high-order 1D NSE soliton. There is a compression of the excessively large initial duration of the wave packet by 3-4 times due to the nonlinear superposition of the structures. The resulting wave structure can be quite effectively captured in the central core. The discreteness of the system is strongly manifested in the second stage $6 \lesssim z \lesssim 20$, when the compressed laser pulse is captured in the central core (Fig.~\ref{ris:ris14}\textbf{(a)}). This capture is clearly seen in Figure~\ref{ris:ris14}{\bf(c)}, which shows the evolution of the compression ratio and the laser pulse energy in the central core. It can be seen that the duration of the wave packet decreased additionally by almost 10 times and the laser pulse energy fraction in the central core $\eta=\int |a|^2d\tau/W$ increased from 0.3 to 0.6. Moreover, the process of the wave packet capturing in the central core is accompanied by a significant dispersion energy loss, which occurs in the first 2-3 oscillations of the field between the center and the ring. At the third stage, the background radiation remaining from the high order NSE soliton escapes from the main pulse gradually. As a result, initial laser pulse \eqref{eq:W0_3} is compressed by $Q = 32$ times with the energy efficiency $\eta = 0.55$ in the formed light bullet (Fig.~\ref{ris:ris14}\textbf{(c)}). At the same time, the evaluation of compression ratio \eqref{eq:32} gives the value $Q \approx 36$ with the energy efficiency $\eta_\text{est} \approx 0.63$ (dotted line in Fig.~\ref{ris:ris14}\textbf{(c)}), i.e. the values are quite close to those obtained in the numerical simulation.

\section{Optimization of initial parameters} \label{sec:8}

\begin{figure}
	\includegraphics[width = 1\linewidth]{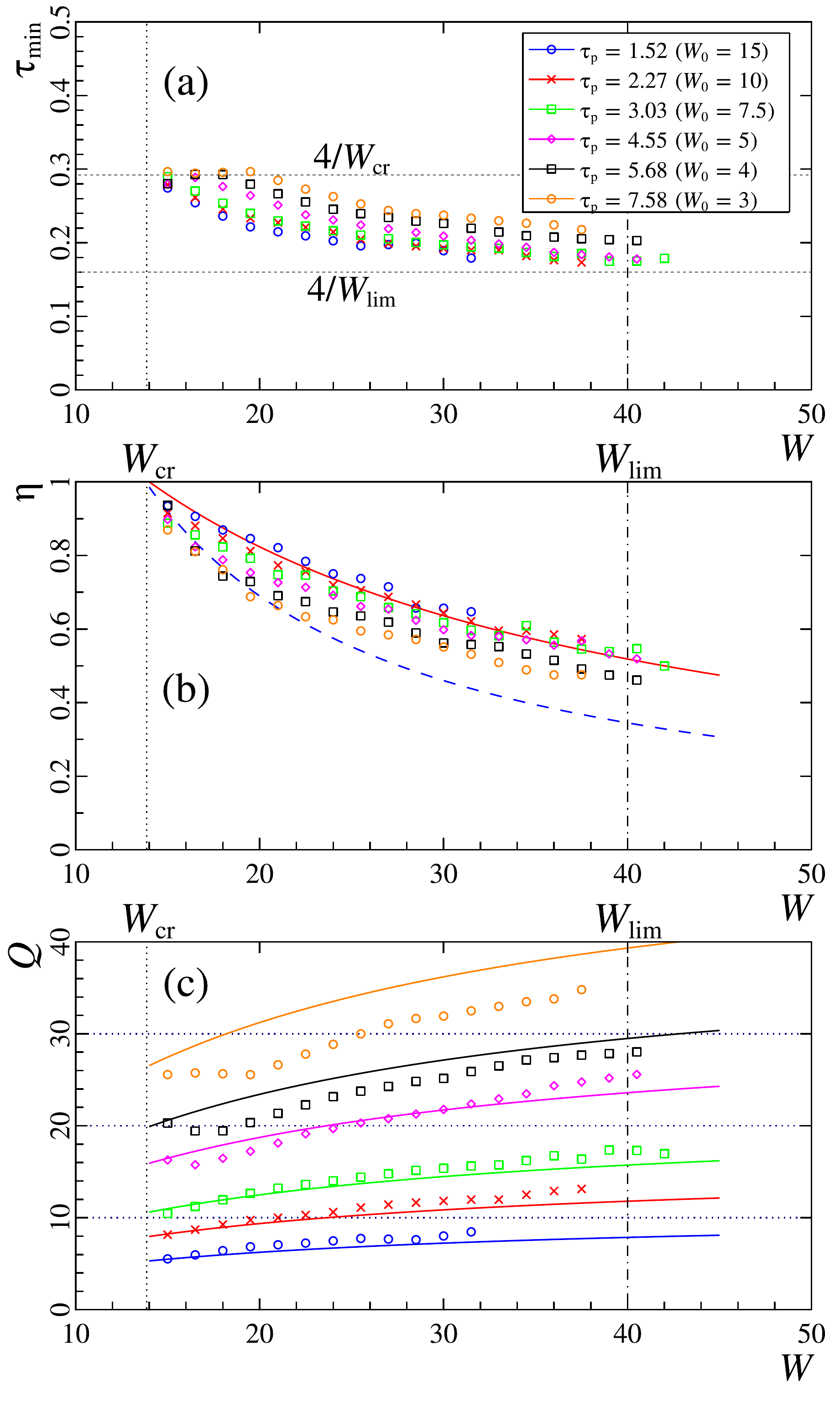}
	\caption{(Color online) Dependencies of the minimum laser pulse duration {\bf (a)}, compression efficiency $\eta=\int |a|^2d\tau/W $ {\bf (b)}, and compression ratio of the wave packet $Q=\tau_p/\tau_p^\textrm{out}$ \textbf{(с)} on the energy $W$ for different initial durations $\tau_p$ (or energies $W_0 \propto 1/\tau_p$). The calculations were performed for a 7-core MCF ($N = 3$, $\chi = 1$). The dotted line in Fig.~\textbf{(a)} shows the duration of the NSE soliton with the energies $W = W_\text{cr}$ and $W = W_\text{lim} \approx 26$. The curves in  Fig.~\textbf{(b)} correspond to approximations \eqref{eq:eta_appr} (solid line) and \eqref{eq:eta_min} (dotted line). The vertical dash-dotted line corresponds to boundary \eqref{eq:w_bound}. The curves in Fig.~\textbf{(c)} shows approximation \eqref{eq:32}.
	} \label{ris:ris11}
\end{figure}

For compression optimization relative to the initial parameters of the laser pulse, integral dependencies of the output parameters are of interest. Figure~\ref{ris:ris11} shows the dependencies of the minimum laser pulse duration $\tau_\text{min}$, efficiency $\eta=\int |a|^2d\tau/W$ and the wave packet compression ratio $Q=\tau_p/\tau_p^\textrm{out}$ on the total energy $W$ for different values of the energy $W_0$ (or the duration of the injected wave packet $\tau_p$). The vertical dashed line in the figure shows the boundary, where the total energy coincides with the critical value, $W = W_\textrm {cr}$. The dots show only the numerically found solutions without splitting, i.e. Fig.~\ref{ris:ris11} contains only solutions leading to the formation of a single light bullet.

Figure \ref{ris:ris11}\textbf{(a)} shows the presence of the minimal duration of the output laser pulse. Moreover, the value of the minimal duration corresponds exactly to the estimates obtained earlier \eqref{eq:wmax_N3}. The main difference from soliton solutions occurs for wave packets with large initial durations and small amplitudes, corresponding to small energies $W_0 \ll W_\text{cr}$. Such long pulses require a significantly larger trace to reduce the duration to values of order of \eqref{eq:17} for $W = W_\text{cr}$ under quasi-one-dimensional compression of the multi-soliton solution. Moreover, at low energies $W_0 = 3$ and 4, the duration of the compressed pulse remains at the level of the critical value at the hysteresis $\tau_\text{cr} = 4/W_\text{cr}$ in a relatively wide energy range $W_\text{cr} <W <24$. Calculations show that an even greater increase in the initial duration of the wave packet (up to $\tau_p \sim 11$ for $W_0 = 2$ and more) results in the impossibility of capturing laser radiation in a single pulse.

The markers in Figure \ref{ris:ris14} show only the results of calculations, in which no pulse splitting is occurred. It can be seen that almost all the markers lie to the left of the analytically predicted boundary \eqref{eq:w_bound}. The reason for the laser pulse splitting can be explained at a qualitative level. At the energy $W \lesssim 30$, the characteristic length of the energy transfer from the cores located in the ring to the central one is commensurate with the dispersion length ($L_\text{dis} \propto 1/\tau_p^2$). Therefore, a smooth amplitude increase in the central core on the scale of the dispersion length is accompanied by an adiabatic decrease in the wave packet duration. However, with an increase in the laser pulse energy $W$, the rapid increase in the amplitude in the central core (the field amplitude increased significantly without changing the duration of the wave packet) can no longer be compensated by the media dispersion. This leads to the development of modulation instability of the wave field in the longitudinal direction. As a result, the laser pulse in the central core is divided into wave structures with the energy $W \sim W_\text{cr}$, on which the dispersion and nonlinear lengths will be approximately the same.

Figure \ref{ris:ris11}{\bf(b)} demonstrates the wave field capture in the central core at energies above the critical value ($W \gtrsim W_\text{cr}$), as in the case of the found distributed soliton \eqref{eq:13}, \eqref{eq:17}. The energy ratio in the central core varies in the range from 1 to 0.5 and coincides well with estimate \eqref{eq:eta_appr} for not-too-long initial pulses with $\tau_p <5$ (or $W_0 \ge 5$, see Fig.~\ref{ris:ris11}). For longer pulses, the efficiency decreases, but it cannot be less than \eqref{eq:eta_min}, which corresponds to the formation of a light bullet with an energy equal to the critical value $W_\text{cr}$.

Figure~\ref{ris:ris11}{\bf(c)} shows the compression ratio $Q$ of the laser pulse depending on the total energy $W$ for different values of the parameter $W_0$. The figure shows good agreement with approximation \eqref{eq:32}. In this case, significant laser pulse compression is achieved for longer laser pulses with a fixed value of the total energy $W>W_\text{cr}$. However, the minimum pulse duration $\tau_\text{min}$ is larger than that for laser pulses with optimal duration \eqref{eq:17}.

So, the results of numerical simulation are in good agreement with the rough analytical estimates \eqref{eq:w_bound}~---~\eqref{eq:32_s} and \eqref{eq:32}. Thus, the use of this method of laser pulses compression in MCF makes it possible to achieve a significant (30-40 times) shortening of the laser pulse duration with the energy efficiency of more than 50\%. At this compressed pulse have no pedestal and no frequency chirp. Note, such a high compression ratio can be achieved in a single fiber by using high order solitons \cite{Agrawal2013}. However, the output pulse will have a large pedestal (up to 80\% of the total pulse energy) and frequency chirp, and the method itself will be sensitive to the length of the fiber. At the same time, the use of long pulses with an energy close to the critical one $W_\text{cr} = 8\sqrt{3}$ allows achieving almost 100\% of energy efficiency at the expense of not more than a twofold decrease in the compression ratio (about 20-25 times).

Let us return to dimensional values \eqref{eq:2}. The soliton energy for duration $\tau_\text{sol}$ must be larger than the critical one: $W = 2|\beta|/\gamma \tau_\text{sol} > W_\text{cr} = 8 \sqrt{3|\beta|\chi_{n,n+1}/(2\gamma^2)}$, where $\beta = \frac{\partial^2k}{\partial\omega^2}$, $\tau_\text{sol} \approx \sqrt{3} \tau_\text{FWHM}$ is the output pulse duration ($\tau_\text{FWHM}$ is full width at half maximum for intensity). This can be done for the coupling coefficient $\chi_{n,n+1} \leq |\beta|/(24\tau_\textrm{sol}^2)$ that is the quite reasonable requirements for the MCF properties. For example, an output pulse with a duration of 50~fs and an energy of 1.4~nJ can be obtained by compressing a soliton-like wave packet with a duration of 300~fs and the same energy using a two-meter quartz MCF with a coupling coefficient $\chi_{n,n+1} \lesssim 1$~m$^{-1}$. The use of longer pulses can significantly increase the compression ratio. For example, the same output pulse (50~fs and 1.4~nJ) can be obtained by compressing a wave packet with duration of 1.5~ps and energy of 2.8~nJ over a length of 10~m of the same MCF light-guide. In principle, the method allows one to obtain shorter pulses, but checking this requires the use of a more general system of equations (which includes high-order dispersion, non-stationary media response and dependence of the group velocity on the field amplitude).

\section{Conclusions}\label{sec:9}

In this paper, we studied propagation of laser pulses in an MCF, which consisted of a central core and an even number $2N$ of cores located in a ring around it. The main attention was paid to the uniform distribution of the wave field around the ring, which interacts most effectively with the field in the central core. Approximate quasi-soliton solutions of the wave field in the considered MCF are found. It is shown that there are three branches of solutions. Branch I corresponds to the in-phase propagation of solitons in cores, and branch II is antiphase (the field in the center is in antiphase to the field on the ring). At low energies $W$, the wave field, corresponding to branch I, is distributed quasi-uniformly over all cores, and at high energies $W$ it is concentrated mainly in the central core. The transition between the two types of distributions is very sharp and occurs at the energy $W_\text{cr} \approx 8 \sqrt{3}$ (Fig.~\ref{ris:ris4}\textbf{(e)}). At the same time, this energy is almost independent of the number of optical cores and the coefficient of coupling with the central core. At high energies, branch II tends to a uniform intensity distribution. Branch III describes a situation in which the wave field is predominantly concentrated only in the ring. The stability analysis showed that only solutions corresponding to branch I are stable. This is confirmed by the results of the numerical simulation.

The found analytical solution for the wave field in branch I has an important feature. At low energies, its wave field is distributed over all MCF cores and has a duration that is many times ($5-6$ times) longer than the duration of the NSE soliton with the same energy. Contrary, almost all of the radiation is concentrated in the central core at high energies, and its duration is close to the duration of the analogous NSE soliton. The form of the solution of branch I at high energies is similar to the light bullet in the MCF. Moreover, the transition from one mode to another is very sharp, through the hysteresis at $W = W_\text{cr}$. This effect can be used to compress laser pulses.

A method for laser pulse compression by light bullet excitation in the MCF is proposed and studied. It is shown that a soliton-like wave packet with an energy greater than the critical value ($W> W_\text{cr}$) is captured into the central core. This process is accompanied by a decrease in the duration of the laser pulse. The formed light bullet has no frequency chirp and no pedestal. We demonstrate analytically and numerically that the compression ratio (the ratio of the initial duration to the output one) as a whole does not depend on the energy and number of optical fibers and is approximately equal to $5.7$. The use of the initial distributions with much longer durations and energies higher than the critical energy allows one to increase the laser pulse compression ratio significantly (up to 30-40 times). Estimates were obtained for the compression efficiency and the minimum attainable duration of the output light bullet, as confirmed by the results of the numerical simulation (Fig.~\ref{ris:ris11}).

This work was supported by the Russian Science Foundation (project No.~16-12-10472).

\end{document}